\documentclass[3p,onecolumn]{elsarticle}
\usepackage{ifpdf}
\pdfoutput=1
\usepackage[usenames]{color}
\usepackage{graphicx}
\usepackage{amssymb}
\usepackage{amsmath}

\journal{Computer Physics Communications}

\begin{document}

\begin{frontmatter}

\title{Fortran and C programs for the time-dependent dipolar
Gross-Pitaevskii equation in an anisotropic trap}

\author[spo]{R.~Kishor~Kumar}
\ead{kishor@if.usp.br}
\author[ift]{Luis~E.~Young-S.}
\ead{luis@ift.unesp.br}
\author[scl-ipb]{Du\v{s}an~Vudragovi\'{c}}
\ead{dusan@ipb.ac.rs}
\author[scl-ipb]{Antun~Bala\v{z}\corref{cor1}}
\ead{antun@ipb.ac.rs}
\author[bdu]{Paulsamy~Muruganandam}
\ead{anand@cnld.bdu.ac.in}
\author[ift]{S.~K.~Adhikari}
\ead{adhikari@ift.unesp.br} 

\address[spo]{Instituto de F\'isica, Universidade de S\~ao Paulo, 05508-090 S\~ao Paulo, SP, Brazil}
\address[ift]{Instituto de F\'{\i}sica Te\'{o}rica, UNESP -- Universidade Estadual Paulista, 01.140-70 S\~{a}o  Paulo, S\~{a}o Paulo, Brazil}
\address[scl-ipb]{Scientific Computing Laboratory, Institute of Physics Belgrade, University of Belgrade, Pregrevica 118, 11080 Belgrade, Serbia}
\address[bdu]{School of Physics, Bharathidasan University, Palkalaiperur Campus, Tiruchirappalli -- 620024, Tamil Nadu, India}

\begin{abstract}

Many of the static and dynamic properties of an atomic Bose-Einstein condensate (BEC) are usually studied by solving the mean-field Gross-Pitaevskii (GP) equation, which is a nonlinear partial differential equation for short-range atomic interaction. More recently, BEC of atoms with 
long-range dipolar atomic interaction are used in theoretical and experimental studies.  
For dipolar atomic interaction, the GP equation is a partial integro-differential equation, requiring complex algorithm for its numerical solution. Here we present numerical algorithms 
for both stationary and non-stationary solutions of the full three-dimensional (3D) GP equation for a dipolar BEC, including the contact interaction.  We also consider the simplified one- (1D) and two-dimensional (2D) GP equations satisfied by cigar- and disk-shaped dipolar BECs. We employ the split-step Crank-Nicolson method with real- and imaginary-time propagations, respectively, for the numerical solution of the GP equation for dynamic and static properties of a dipolar BEC. The atoms are considered to be polarized along the $z$ axis and we consider  ten different cases, e.g., stationary and non-stationary solutions of the GP equation for a dipolar BEC in 1D (along $x$ and $z$ axes), 2D (in $x-y$ and $x-z$ planes), and 3D, and we provide working codes in Fortran 90/95 and C for these ten  cases (twenty programs in all).
We  present numerical results for energy, chemical potential, root-mean-square sizes and density of the dipolar BECs and, where available, compare them 
with results of other authors and of variational and Thomas-Fermi approximations.

\end{abstract}

\begin{keyword}
Bose-Einstein condensate; Gross-Pitaevskii equation; Split-step Crank-Nicolson
scheme; Real- and imaginary-time propagation; Fortran and C programs; Dipolar atoms

\PACS 67.85Hj; 03.75.Lm;  03.75.Nt; 64.60.Cn
\end{keyword}
\end{frontmatter}

\newpage

{\noindent {\bf  
Program summary}}
\vskip .3 cm

{\noindent{{\it Program title}: (i) imag1dZ, (ii) imag1dX, (iii) imag2dXY, (iv) imag2dXZ, (v) imag3d, (vi) real1dZ, (vii)  real1dX, (viii) real2dXY, (ix) real2dXZ,
(x) real3d}}

{\noindent{{\it Catalogue identifier}:}} AEWL\_v1\_0

{\noindent{{\it Program summary URL}: http://cpc.cs.qub.ac.uk/summaries/AEWL\_v1\_0.html}}

{\noindent{{\it Program obtainable from}: CPC Program Library, Queen’s University, Belfast, N. Ireland}}

{\noindent{{\it Licensing provisions}: Standard CPC licence, http://cpc.cs.qub.ac.uk/licence/licence.html}}



{\noindent{{\it Distribution format}: tar.gz}}

{\noindent{{\it Programming language}: Fortran 90/95 and C}}

{\noindent{{\it Computer}: Any modern computer with Fortran 90/95 or C language compiler installed.}}

{\noindent{{\it Operating system}: Linux, Unix}}

{\noindent{{\it RAM}: 1 GB (i,ii), 2 GB (iii,iv), 4 GB (v), 2 GB (vi,vii), 4 GB (viii,ix), 8 GB (x)}}

{\noindent{{\it Classification}: 2.9, 4.3, 4.12}}

{\noindent{{\it Nature of problem}: These programs are designed to solve the time-dependent  nonlinear
partial differential Gross-Pitaevskii (GP) equation with contact and dipolar interactions 
in one, two or three space dimensions in a harmonic anisotropic trap. The GP equation
describes the properties of a dilute trapped Bose-Einstein condensate.}}

{\noindent{{\it Solution method}: The time-dependent GP equation is solved by the split-step Crank-Nicolson
method by discretizing in space and time. The discretized equation is then solved by propagation, in
either imaginary or real time, over small time steps. The contribution of the dipolar interaction is evaluated by a Fourier transformation (FT) to momentum space using a convolution theorem. 
The method yields the solution of stationary and/or
non-stationary problems.}}

{\noindent{{\it Additional comments}: This package consists of 12 programs, see “Program title”, above. Fortran 90/95 and C
versions are provided for each of the 6  programs. For the particular purpose of each program please see below.}}

{\noindent{{\it Running time}: Minutes on a medium PC (i, ii), tens of minutes on a medium  PC (iii, iv),
tens of minutes on a good workstation (v, vi).}}

\newpage

\section{Introduction}

After the experimental realization of atomic Bose-Einstein condensate (BEC) of
alkali-metal and some other 
atoms, there has been a great deal of theoretical activity  
in studying the statics and dynamics of the condensate using the mean-field 
time-dependent Gross-Pitaevskii (GP) equation under different trap symmetries \cite{Dalfovo1999}. 
The GP equation in three dimensions (3D) is a nonlinear partial differential equation in 
three space variables and a time variable and its numerical solution is indeed a difficult task specially for large nonlinearities encountered in 
realistic experimental 
situations \cite{mu-ad}. 
Very special numerical algorithms are necessary for its precise 
numerical solution. 
 In the case of alkali-metal atoms the atomic interaction in 
dilute BEC is essentially of short-range in nature and is approximated 
by a contact interaction and at zero tempearture is
parametrized by a single parameter in a dilute BEC $-$ the 
s-wave atomic scattering length. Under this approximation the atomic interaction is represented by a cubic nonlinearity in the GP equation. Recently, we published the Fortran \cite{CPC1}
and C \cite{CPC2} versions  of useful programs for the numerical solution 
of the time-dependent GP equation with cubic nonlinearity under 
different trap symmetries using split-step Crank-Nicolson scheme and real- and imaginary-time propagations.   
Since then, these programs enjoyed widespread use \cite{use}.

More recently, there has been experimental observation of BEC of $^{52}$Cr \cite{cr},
$^{164}$Dy \cite{dy} and $^{168}$Er \cite{er} atoms with large magnetic dipole
moments. In this paper, for all trap symmetries the dipolar atoms are considered to be 
 polarized along the $z$ axis. 
 In these cases
the atomic interaction has a long-range dipolar counterpart in addition to the 
usual contact interaction.  The s-wave contact interaction    is 
local and spherically symmetric, whereas the dipolar interaction acting in all partial waves
is  nonlocal and asymmetric. The resulting GP equation in this case is a partial 
integro-differential equation and  special algorithms are required for its numerical solution.
Different approaches to the numerical solution of the dipolar GP equation have been suggested 
\cite{Bao2010,Yi2001,Goral2002a,dip3,Blakie,dip5}.  Yi and You \cite{Yi2001} solve the dipolar GP equation for 
axially-symmetric trap while they perform the angular integral of the dipolar term, thus 
reducing it to one in axial ($z, z'$) and radial ($\rho, \rho'$) variables involving standard Elliptical integrals. The dipolar term is  regularized by a cut-off at small distances and then evaluated numerically. 
The dipolar GP  equation is then solved by imaginary-time propagation. G\`oral and Santos \cite{Goral2002a} treat the dipolar term by a convolution theorem without approximation, thus transforming it to an inverse Fourier transformation (FT) of a product 
of the FT of the dipolar potential and the condensate density. The FT and inverse FT are then numerically evaluated by standard fast Fourier transformation  (FFT) routines in Cartesian coordinates. The ground state of the system is obtained by employing
a standard split-operator technique in imaginary time.
This approach is used by some others \cite{Parker}. Ronen {\it et al.} perform the angular integral in the dipolar term using 
axial symmetry. To evaluate it, in stead of FT in $x,y,$ and $z$ \cite{Goral2002a}, they use Hankel transformation in  the radial $\rho$ variable and
FT in the axial $z$ variable. The ground state wave function is then obtained by imaginary-time propagation and dynamics by real-time propagation. 
This approach is also used by some others \cite{Blume}. 
Bao {\it et al.} use  Euler sine pseudospectral method for computing the ground states and a time-splitting sine pseudospectral method for computing the dynamics of dipolar BECs \cite{Bao2010}.
 Blakie {\it et al.}  solve the projected dipolar GP equation using a Hermite polynomial-based spectral representation \cite{Blakie}. Lahaye {\it et al.} use FT in $x,y,$ and $z$ to evaluate the dipolar term and 
employ imaginary- and real-time propagation after Crank-Nicolson discretization for stationary and nonstationary solution of the dipolar GP equation \cite{dip5}. 

Here here we provide Fortran and C versions of programs for the solution of the dipolar GP equation in a fully 
anisotropic 3D trap by 
real- and imaginary-time propagation. 
We use split-step Crank-Nicolson scheme for the nondipolar part as in Refs. \cite{CPC1,CPC2} and the dipolar term is 
treated by FT in $x,y,z$ variables.   We also 
present the Fortran and C programs for  reduced dipolar GP equation in 
one (1D) and two dimensions (2D) appropriate for a cigar- and disk-shaped BEC under tight radial ($\rho$) and axial  ($z$)
trapping, respectively \cite{lsa}.  In the 1D case, we consider two possibilities: the 1D BEC could be aligned along the polarization direction $z$ or be aligned perpendicular to the polarization direction along $x$ axis. Similarly, in the 2D case,  two possibilities are considered taking  the 2D plane as $x-y$, perpendicular to polarization direction $z$ or as $x-z$ containing the polarization direction. This amounts to five different trapping possibilities $-$ two in 1D and 2D each and one in 3D $-$ and  two solution schemes involving real- and imaginary-time propagation resulting in ten programs each in Fortran and C.

In Sec. \ref{sec:gpe3D} we present the 3D dipolar GP equation in an anisotropic trap. In addition to presenting the mean-field model and a general 
scheme for its numerical solution  in Secs. 
\ref{model} and \ref{method}, 
we also present two 
approximate solution schemes in 
 Secs. 
\ref{sec:vari} and \ref{sec:tf}, 
$-$ Gaussian variational approximation and Thomas-Fermi (TF)  approximation $-$  in this case. The 
reduced 1D and 2D GP equations appropriate for a cigar- and a disk-shaped dipolar BEC are next presented in Secs. 
\ref{1dgpe} and \ref{2dgpe}, respectively. The details about the computer programs, and their input/output files, etc.  are given in Sec. \ref{programs}. The numerical method and results are given in Sec. \ref{num}. Finally, a 
brief summary is given in Sec. \ref{summary}.

\section{Gross-Pitaevskii (GP) equation for dipolar condensates in three dimensions}

\label{sec:gpe3D}

\subsection{The mean-field Gross-Pitaevskii equation}
 
\label{model}

At ultra-low temperatures the properties of a dipolar condensate of $N_{\mathrm{at}}$ 
atoms, each of mass $m$, can be described by the mean-field 
GP equation with nonlocal nonlinearity of the 
form:~\cite{Yi2001,Santos2000}
\begin{align}
i\hbar\frac{\partial \phi({\mathbf r},t)}{\partial t} = & 
\left[-\frac{\hbar^2}{2m}\nabla^2+V_{\text{trap}}({\mathbf r}) + \frac{4\pi\hbar^2a N_{\mathrm{at}}}{m}\left\vert 
\phi({\mathbf r},t)\right\vert^2  +N_{\mathrm{at}} \int U_{\mathrm{dd}}({\mathbf  r}-{\mathbf r}') 
\left\vert\phi({\mathbf r}',t)\right\vert^2 d{\mathbf r}' 
\right]\phi({\mathbf r},t),
\label{eqn:dgpe}
\end{align}
where $\int d{\bf r}  \vert  \phi({\mathbf r},t) \vert  ^2=1.$ 
 The trapping potential, $V_{\text{trap}}$ is 
assumed to be fully  asymmetric of the form
\begin{align}
V_{\text{trap}}({\mathbf r}) = \frac{1}{2} m \left(\omega_x^2 x^2+
\omega_y^2 y^2+ \omega_z^2 z^2 \right) \notag
\end{align}
where $\omega_x, \omega_y $ and  
$\omega_z$ are the trap frequencies, $a$ the atomic scattering length.  
The dipolar interaction, for magnetic dipoles, is given by \cite{Goral2002a,Blume} 
\begin{align}
U_{\mathrm{dd}}(\bf R)=\frac{\mu_0 \bar \mu^2}{4\pi}\frac{1-3\cos^2 \theta}{ \vert  {\bf R} \vert  ^3},
\end{align}
where ${\bf R= r -r'}$ determines the relative position of dipoles and $\theta$ 
is the angle between ${\bf R}$ and the direction of polarization $z$, 
 $\mu_0$ is the permeability of free space 
and $\bar \mu$ is the dipole moment of the condensate atom.
To compare the 
contact and dipolar interactions, often it is useful to introduce the 
length scale $a_{\mathrm{dd}}\equiv \mu_0 \bar \mu^2  m/(12\pi \hbar^2)$ \cite{cr}.

Convenient dimensionless parameters can be defined in terms of a reference
 frequency $\bar \omega $ and the corresponding
oscillator length $l=\sqrt{\hbar/(m\bar \omega)}$.
Using dimensionless variables ${\bar {\bf r}}= {\bf r}/l,
{\bar {\bf R}}= {\bf R}/l,
 \bar a = a/l, \bar 
a_{\mathrm{dd}}= a_{\mathrm{dd}}/l, \bar t = 
t\bar \omega$, $\bar x = x/l, \bar y = y/l, \bar z = z/l, \bar \phi 
= l^{3/2}\phi$, 
  Eq. 
(\ref{eqn:dgpe})  can be rewritten (after removing the overhead bar from all the variables) as 
\begin{align}\label{gpe3d}
i \frac{\partial   \phi( {\mathbf { r}},{  t})
}{\partial   t}
 & = \biggr[ -\frac{1}{2}\nabla^2+ \frac{1}{2}
\left(  {\gamma^2  x^2+
\nu^2  y^2}
+\lambda^{2}  z^2\right) +{4 \pi  a N_{\mathrm{at}}} \vert  {  \phi} \vert  ^2 
+3 N_{\mathrm{at}}  a_{\mathrm{dd}}\int   V_{\mathrm{dd}}^{3D}({\bf { R}})
 \vert     \phi( {\mathbf { r}}',  t) \vert  ^2 d {\mathbf { r}}'
\biggr]  \phi( {\mathbf { r}},{ t}),
\end{align}
with 
\begin{align}
  V_{\mathrm{dd}}^{3D}  ({\bf { R}}) =  \frac{1-3\cos^2 \theta}
{ \vert  {\bf { R}} \vert  ^3}\, ,
\end{align}
where 
$\gamma= \omega_x/\bar \omega, \nu=\omega_y/\bar \omega,
\lambda =\omega_z/\bar \omega$. The reference frequency $\bar \omega $ can be taken as one of the frequencies $\omega_x, \omega_y$ 
or $\omega_z$ or their geometric mean $(\omega_x \omega_y \omega_z)^{1/3}$. 
In the following we shall use Eq. (\ref{gpe3d}) where we have  removed the 
`bar' from all variables.

Although we are mostly interested in the numerical solution of Eq. (\ref{gpe3d}), in the following we 
describe two {\it analytical} approximation methods for its solution in the axially-symmetric case. These approximation methods $-$ the Gaussian variational 
 and  TF approximations $-$ provide reasonably accurate results under some limiting conditions 
and will be used for comparison with the numerical results. Also, we present reduced 1D and 2D mean-field GP equations appropriate for the description of a cigar and disk-shaped dipolar BEC under appropriate trapping condition. The numerical solution and variational approximation
of these reduced equations will be discussed in this paper. 
A brief algebraic description of these topics are presented for the sake of completeness as appropriate for this study. For a full description of the same the reader is referred to the original publications.

\subsection{Methodology}

\label{method}

We perform numerical simulation of the 3D GP equation
(\ref{gpe3d}) using the split-step Crank-Nicolson method described in detail
in Ref. \cite{CPC1}. Here we present the procedure to include the dipolar 
term in that algorithm.
The inclusion of the dipolar integral term in the GP equation
in coordinate space is not straightforward due to the
singular behavior of the dipolar potential  at short distances. 
It is interesting to note that 
this integral is well defined and finite. 
This problem  has been
tackled by evaluating the dipolar term in the momentum
(k) space, where we do not face a singular behavior. The integral can be simplified in Fourier space
by means of convolution as
\begin{align}\label{con}
\int d{\bf r}'V_{\mathrm{dd}}^{3D}({\bf r} - {\bf r}') n({\bf r'},t)  = \int 
\frac{d{\bf k}}{(2\pi)^3}e^{-i {\bf k\cdot r}}
\widetilde V_{\mathrm{dd}}^{3D}({\bf k})\widetilde n ({\bf k},t),
\end{align}
where $n({\bf r},t)= \vert  \phi({\bf r},t) \vert  ^2$. The Fourier transformation (FT) and
inverse FT, respectively, are defined by 
\begin{align}
\widetilde A({\bf k})=\int d {\bf r}A({\bf r})e^
{i {\bf k}\cdot {\bf r}},\quad
A({\bf r})=\frac{1}{(2\pi)^3}\int d {\bf k}
\widetilde A({\bf k})e^ {-i {\bf k}\cdot {\bf r}}.
\end{align}
The FT of the dipole potential
can be obtained analytically 
\cite{prog}
\begin{align}\label{pqr}
\widetilde V_{\mathrm{dd}}^{3D}({\bf k}) \equiv \frac{4\pi}{3} h_{3D}({\bf k})
=\frac{4\pi}{3} \left(\frac{3k_z^2}{{\bf k}^2}-1  \right),
\end{align}
so that
\begin{align} \label{xyft}
\int d{\bf r}'V_{\mathrm{dd}}^{3D}({\bf r} - {\bf r}') n({\bf r'},t)  = \frac{4\pi}{3}\int 
\frac{d{\bf k}}{(2\pi)^3}e^{-i {\bf k\cdot r}}
 h_{3D}({\bf k})
\widetilde n ({\bf k},t).
\end{align}
To obtain Eq. (\ref{pqr}), first the angular integration is performed. 
Then  
a cut-off at small $r$ is introduced to perform the radial integration
and eventually  the zero cut-off limit is taken in the final result as shown in Appendix A of 
Ref.  \cite{prog}.  
The FT of density $ n({\bf r})  $ is evaluated
numerically by means of a standard  FFT 
algorithm. The dipolar integral in Eq. (\ref{gpe3d}) involving the
FT of density multiplied by FT of dipolar interaction is
evaluated by the convolution theorem (\ref{con}). The inverse
FT is taken by means of the standard FFT algorithm. The
FFT algorithm is carried out in Cartesian coordinates
and hence the GP equation is solved in 3D
irrespective of the symmetry of the trapping potential.
The dipolar interaction integrals in 1D and 2D GP equations 
are also evaluated in momentum spaces.  
The solution algorithm of the GP equation by the split-step 
Crank-Nicolson method is adopted from Refs. \cite{CPC1,CPC2}. 

{
The 3D GP equation (\ref{gpe3d}) is numerically the most difficult  to solve 
involving large RAM and CPU time.  A requirement for the success of the split-step 
Crank-Nicolson method using a FT continuous at the origin 
is that on the boundary of the space discretization region 
the wave function and the interaction term should vanish. For the long-range  
dipolar potential this is not true and the FT (\ref{pqr}) is discontinuous at the origin. 
 The space domain (from $-\infty$ to $+\infty$) cannot be restricted to a small region in space just covering the spatial extention of the BEC as the same domain
is also used to calculate the FT and inverse FT used in treating the long-range  dipolar potential. The use and success of FFT implies a set of noninteracting 3D periodic lattice of BECs in different unit cells. This is not true for long-range dipolar interaction which will lead to an
 interaction between BECs in different cells. Thus, boundary effects can play a role when finding the FT. Hence a suffciently large space domain is to be used to have accurate values of the FT involving the long-range dipolar potential.  It was suggested \cite{dip3} that
this could be avoided by truncating the dipolar interaction conveniently  at large distances  $r=R$
 so that it does not affect the boundary, provided $R$ is taken to be larger than the size of the condensate. 
Then the truncated dipolar potential will cover  
the whole condensate wave function and will have a  continuous FT 
at the origin. This will improve the 
accuracy of a  calculation using a small space domain. The  FT of the dipolar potential truncated at $r=R$,
as suggested in Ref.  
\cite{dip3},  is used in the numerical  routines
\begin{align}\label{cutoff}
\widetilde V_{\mathrm{dd}}^{3D}({\bf k})  
=\frac{4\pi}{3} \left(\frac{3k_z^2}{{\bf k}^2}-1  \right)\left[1+3 \frac{\cos(kR)}{k^2R^2}
-3\frac{\sin(kR)}{k^3R^3} \right],  \quad \quad k=|{\bf k}|.
\end{align} 
Needless to say, the diffculty in using a large space domain is the most severe in 3D. In 3D programs the cut-off $R$  of Eq. (9) improves the accuracy of calculation and a smaller space region can be used in numerical treatment. In 1D and 2D, a larger space domain can be used relatively easily and no cut-off has been used. Also, no convenient and effcient analytic cut-off is known in 1D and 2D \cite{dip3}.
The truncated dipolar potential (\ref{cutoff}) has only been used in the numerical programs in 3D, e.g.,
imag3d* and real3d*. In all other numerical programs in 1D and 2D, 
and in all analytic results reported in the following the untruncated potential (\ref{pqr}) has been used. 
}

\subsection{Gaussian variational approximation}

\label{sec:vari}

In the axially-symmetric case ($\gamma=\nu$), convenient analytic
Lagrangian  variational approximation  of Eq. (\ref{gpe3d}) can be obtained with the following Gaussian
ansatz for the wave function \cite{gv}
\begin{align}
\phi({\bf r},t)=\frac{\pi^{-3/4}}{w_\rho(t)\sqrt{w_z(t)}}
\exp\left[-\frac{\rho^2}
{2w_\rho^2(t)}-\frac{z^2}{2w_z^2(t)}+i\alpha(t)\rho^2+i\beta(t) z^2\right]
\end{align}
where ${\bf r}=\{\boldsymbol \rho,z   \}, {\boldsymbol \rho}=\{x,y\}$, $w_\rho(t)$ 
and $w_z(t)$ are widths and $\alpha(t)$ and 
$\beta(t)$ are chirps. The time dependence of the variational parameters 
$w_\rho(t),$  $w_z(t)$, $\alpha(t)$ and 
$\beta(t)$ will not be explicitly shown in the following.

The Lagrangian density corresponding to  Eq. (\ref{gpe3d}) is given by 
\begin{align} \label{lagden}
{\cal L}({\bf r})&= \frac{i}{2}\left[ \phi({\bf r})\frac{\phi ^*({\bf r})}{\partial t} -\phi^*({\bf r})  \frac{\partial \phi({\bf r})}{\partial t} \right]  + \frac{|\nabla \phi({\bf r})|^2}{2}+ 
\frac{1}{2}(\gamma^2\rho^2+\lambda^2 z^2)|\phi({\bf r})|^2+
 2 \pi a N_{\mathrm{at}} |\phi({\bf r})|^4 \nonumber \\
&+ \frac{3a_{\mathrm{dd}}N_{\mathrm{at}}}{2}
|\phi({\bf r})|^2\int V_{\mathrm{dd}}({\bf R})|\phi({\bf r'})|^2 d {\bf r'}.
 \end{align}
Consequently, the effective Lagrangian $L \equiv \int {\cal L}({\bf r})
d{\bf r}$ (per particle) becomes \cite{cr,joptb}
\begin{align} \label{efflag}
L=   \omega_\rho^2\dot  \alpha+\frac{ \omega_z^2\dot  \beta }{2}+ \frac{\gamma^2\omega_\rho^2}{2}+\frac{\lambda^2\omega_z^2}{4} + \frac{1}{2\omega_\rho^2}
+\frac{1}{4\omega_z^2}+ 2 \omega_\rho^2\alpha^2+\omega_z^2 \beta^2
+\frac{N_{\mathrm{at}}[a-a_{\mathrm{dd}}f(\kappa)]}{\sqrt{2\pi}\omega_\rho^2\omega_z}.
 \end{align}
The Euler-Lagrangian equations with this Lagrangian  leads to the 
following set of coupled ordinary differential equations (ODE) for the widths $w_\rho$ and $w_z$ \cite{laserphysics}
 \begin{align} &
\ddot{w}_{\rho}
+\gamma^2 w_\rho
=
\frac{1}{w_\rho^3} +\frac{
N_{\mathrm{at}}}{\sqrt{2\pi}} \frac{  \left[2{a} - a_{\mathrm{dd}}
{g(\kappa) }\right]  }{w_\rho^3w_{z}}
,
\label{f1} \\ & \ddot{w}_{z}
 +\lambda^2 w_z =
\frac{1}{w_z^3}+ \frac{ 2N_{\mathrm{at}}}{\sqrt{2\pi}}
\frac{ \left[{a}-a_{\mathrm{dd}}
c(\kappa)\right]  }{w_\rho^2w_z^2} , \label{f2}
\end{align}
with $\kappa=w_\rho/w_z$ and 
\begin{align}
& g(\kappa)=\frac{2-7\kappa^2-4\kappa^4+9\kappa^4 d(\kappa)}{(1-\kappa^2)^2}, \\
& c(\kappa) =\frac{1+10\kappa^2 -2\kappa^4 -9\kappa^2 d(\kappa)}{(1-\kappa^2)^2},\\
\label{fkappa}
&f(\kappa)= \frac{1+2\kappa^2-3\kappa^2d(\kappa)}{1-\kappa^2}, \quad d(\kappa)= \frac{\mbox{atanh}\sqrt{1-\kappa^2}}{\sqrt{1-\kappa^2}}.
\end{align}
The widths of a (time-independent) stationary state  are obtained from Eqs. 
(\ref{f1}) and (\ref{f2}) by setting  $\ddot{w}_{\rho}=\ddot{w}_{z}=0$.
The energy (per particle) of the stationary state is the Lagrangian 
(\ref{efflag}) with $ \alpha
= \beta =0$, e.g.,
\begin{align}\label{ve3d}
\frac{E}{N_{\mathrm{at}}} =  &\, \frac{1}{2w_\rho^2}+
\frac{1}{4w_z^2}+\frac{N_{\mathrm{at}}[a-a_{\mathrm{dd}}f(\kappa)]}{\sqrt{2\pi} w_zw_\rho^2}
+\frac{\gamma^2 w_\rho^2}{2}
+\frac{\lambda^2 w_z^2}{4}.
\end{align}
The chemical potential $\mu =\partial E/\partial N_{\mathrm{at}}$ of the  stationary state  is given
 by \cite{laserphysics}
\begin{align}\label{vmu3d}
\mu =  &\, \frac{1}{2w_\rho^2}+
\frac{1}{4w_z^2}+\frac{2N_{\mathrm{at}}[a-a_{\mathrm{dd}}f(\kappa)]}{\sqrt{2\pi} w_zw_\rho^2}
+\frac{\gamma^2 w_\rho^2}{2}
+\frac{\lambda^2 w_z^2}{4}.
\end{align}

\subsection{Thomas-Fermi (TF) approximation}

\label{sec:tf}

In the  time-dependent axially-symmetric GP equation (\ref{gpe3d}), when the atomic interaction 
term is large compared to the kinetic energy gradient term, the kinetic energy 
can be neglected and the useful TF approximation emerges.  
We assume the normalized  density of the dipolar BEC  of the form
\cite{Dalfovo1999,ODell2004,Eberlein2005,Parker2008}
\begin{align}
 n({\mathbf r},t)\equiv |\phi({\bf r},t)|^2 = \frac{15}{8\pi R_\rho^2(t) R_z(t)}
\left[1-\frac{\rho^2}{R_\rho^2(t)}-
\frac{z^2}{R_z^2(t)}\right],
\label{tfden}
\end{align}
where $R_\rho(t)$ and $R_z(t)$ are the radial and axial sizes. The time dependence of these 
sizes will not be explicitly shown in the following.
Using the parabolic density (\ref{tfden}), 
the energy functional $E_{TF}$ may be written as~\cite{Eberlein2005}
\begin{align}
E_{TF}\equiv E_{\mathrm{trap}}+E_{\mathrm{int}} = \biggr[\frac{N(2\gamma^2 R_\rho^2 +\lambda^{2}  R_z^2)}{14}\biggr] + \biggr[
\frac{15}{28\pi } \frac{  4\pi aN_{\mathrm{at}}^2}{R_\rho^2 R_z} \left\{ 
1 -\frac{a_{\mathrm{dd}}}{a} {f(\bar \kappa)} \right\}\biggr], \label{tfen}
\end{align}
where $\bar \kappa = R_\rho/R_z$ is the ratio of the condensate 
sizes and $f(\bar\kappa)$ is given by Eq. (\ref{fkappa}).
In Eq. (\ref{tfen}),  $E_{\mathrm{trap}}$ is the energy in the trap and $E_{\mathrm{int}}$ is the interaction or release energy in the TF approximation. 
In the TF regime one has the following 
set of coupled  ODEs for the evolution of the condensate sizes
\cite{ODell2004}:  
\begin{align}
  \ddot R_\rho & = 
-{R_\rho}\gamma^2+\frac{15aN_{\mathrm{at}}}{R_\rho R_z 
}\left[\frac{1}{R_\rho^2}-\frac{a_{\mathrm{dd}}}{a}\left(\frac{1}
{R_\rho^2}+\frac{3}{2}\frac{f(\bar\kappa)}{R_\rho^2
-R_z^2}\right)\right]\label{eq:tf_dyn:a} 
,\\ \ddot R_z & = -\lambda^{2} R_z+\frac{15aN_{\mathrm{at}}}{R_\rho^2 
}\left[\frac{1}{R_z^2}+\frac{2a_{\mathrm{dd}}}{a}\left(\frac{1}{R_z^2}+
\frac{3}{2}\frac{f(\bar\kappa)}{R_\rho^2-R_z^2}\right)\right].
\label{eq:tf_dyn:b}. 
\end{align} 
The  sizes of a stationary state can be calculated from Eqs. (\ref{eq:tf_dyn:a})
and (\ref{eq:tf_dyn:b}) by setting the time derivatives 
$\ddot R_\rho$ and $\ddot R_z$ to zero leading to the transcendental equation 
for $\bar \kappa$ \cite{ODell2004}
\begin{align}\label{ktf}
3\bar \kappa^2 \frac{a_{\mathrm{dd}}}{a}\left[ \left(1
+\frac{\lambda^2}{2\gamma^2}\right)\frac{f(\bar \kappa)}
{1-\bar \kappa^2}-1\right]+\left(\frac{a_{\mathrm{dd}}}{a}-1\right)\left(\bar\kappa^2-\frac{\lambda^2}{\gamma^2}\right)=0,
\end{align}
and 
\begin{align}\label{rtf}
R_\rho=\left[ 
\frac{15aN_{\mathrm{at}}\bar\kappa}{\gamma^{2}}
\left\{1+\frac{a_{\mathrm{dd}}}{a}
\left(\frac{3}{2}\frac{\bar\kappa^2f(\bar\kappa)}
{1-\bar\kappa^2}-1\right)\right\}
\right]^{1/5},
\end{align}
with 
 $R_z=R_\rho/\bar\kappa$. The chemical potential is given by 
\cite{Eberlein2005} 
\begin{align}\label{tfch}
\mu_{TF}
\equiv E_{\mathrm{trap}}+2E_{\mathrm{int}}
=\frac{15}{8\pi}\frac{4\pi a N_{\mathrm{at}}}{ R_\rho^2 R_z}\left[1-\frac{a_{\mathrm{dd}}}{a}f(\bar \kappa)\right].
\end{align}  
We have the identities $ E_{TF}/N_{\mathrm{at}}=5\mu_{TF}/7,  E_{\mathrm{int}}/N_{\mathrm{at}}=2\mu_{TF}/7,
 E_{\mathrm{trap}}/N_{\mathrm{at}}=3\mu_{TF}/7.$

\subsection{One-dimensional GP equation for a cigar-shaped dipolar BEC }

\label{1dgpe}

\subsubsection{$z$ direction}

For a cigar-shaped dipolar BEC with a strong axially-symmetric ($\nu=\gamma$)
radial
trap ($\lambda <\nu$, $\gamma$), 
we assume that the dynamics of the BEC 
in the radial direction is confined in the radial ground state \cite{laserphysics,santos1d,1d-dell}
\begin{align}
\phi(\boldsymbol \rho) = \exp(-\rho^2/2d_\rho^2)/
(d_\rho\sqrt\pi), \quad \gamma d_\rho^2
=1,\quad \boldsymbol \rho \equiv (x,y),  
\end{align}
of the transverse trap and the wave 
function $\phi({\bf r})$ can be written as 
\begin{align}\label{an1}
\phi({\bf r},t) \equiv 
\phi_{1D}(z,t) \times \phi(\boldsymbol \rho) =\frac{1}{\sqrt{\pi d_\rho^2}}\exp\left[  -\frac
{\rho^2}{2d_\rho^2}\right]  \phi_{1D}(z,t),
\end{align}
where $\phi_{1D}(z,t)$ is the effective 1D wave function for the axial dynamics and 
$d_\rho$ is the radial harmonic oscillator length.

To derive the
effective 1D equation for the cigar-shaped dipolar BEC,
we substitute the ansatz (\ref{an1}) in Eq. (\ref{gpe3d}), multiply by the
ground-state wave function $\phi(\boldsymbol \rho)$ and integrate over $\boldsymbol \rho$ to get
the 1D equation \cite{laserphysics,santos1d}
\begin{align}  \label{gpe1d}
&
i\frac{\partial \phi_{1D}(z,t)}{\partial t}=\left[-\frac{\partial_z^2}{2}
+\frac{\lambda^2 z^2}{2}+\frac{2 aN_{\mathrm{at}} \vert  \phi_{1D} \vert  ^2}{ d_\rho^2}
+{3a_{\mathrm{dd}}N_{\mathrm{at}}}\int_{-\infty}^{\infty} V^{1D}_{\mathrm{dd}}( \vert  z- z' \vert  ) \vert  \phi_{1D}(z',t) \vert  ^2
dz'\right] \phi_{1D}(z,t),\\
& V_{\mathrm{dd}}^{1D}(Z)=\frac{2\pi }{\sqrt 2 d_\rho}
\left[\frac{4}{3}\delta(\sqrt w)+2\sqrt w-\sqrt \pi(1+2w)e^w
\{1-\text{erf}(\sqrt w)\}\right],
\end{align}
where $w=[Z/(\sqrt 2 d_\rho)]^2$, $Z= \vert  z-z' \vert$. Here and in all reductions in Secs. 
\ref{1dgpe} and \ref{2dgpe} we use the untruncated dipolar potential (\ref{pqr}) and not the truncated
potential (\ref{cutoff}). 
The integral term in the 
1D GP equation (\ref{gpe1d}) is conveniently evaluated in momentum space 
using the   following  convolution identity \cite{laserphysics}
\begin{align}\label{iden}
\int_{-\infty}^{\infty} V^{1D}_{\mathrm{dd}}( \vert  z- z' \vert  ) \vert  \phi_{1D}(z',t) \vert  ^2
dz'= \frac{4\pi}{3}\int_{-\infty}^{\infty}\frac{dk_z}{2\pi}e^{-ik_z z}\widetilde 
n (k_z,t)h_{1D}\left(\frac{k_zd_\rho}{\sqrt 2}   \right),
\end{align}
where   
\begin{align}&
\widetilde n(k_z,t)= \int_{-\infty}^{\infty} e^{ik_z z} \vert  \phi_{1D}(z,t) \vert  ^2dz,\\
& \widetilde n({\bf k}_\rho)= \int e^{i{\bf k}_\rho \cdot  \rho} \vert  \phi_{2D}(\rho) \vert  ^2d\rho= e^{-k_\rho^2  d_\rho^2/4}, \quad  k_\rho=\sqrt{k_x^2+k_y^2} \\ &
h_{1D}(\zeta)\equiv\frac{1}{(2\pi)^2} \int d{\bf k}_\rho   \left[ \frac{3k_z^2}{{\bf k}^2} -1          \right]   |\widetilde n({\bf k}_\rho)|^2       
=\frac{1}{2\pi d_\rho^2}
\int_0^\infty du \left[\frac{3\zeta^2}{u+\zeta^2}   -1\right]
e^{-u}, \quad \zeta= \frac{k_zd_\rho}{\sqrt 2} .
\end{align}

The 1D GP equation (\ref{gpe1d}) can be solved analytically using the 
 Lagrangian variational formalism with the following Gaussian
ansatz for the 
wave function \cite{laserphysics}:
\begin{align}
 \phi_{1D}(z,t)=\frac{\pi^{-1/4}}{\sqrt{w_z(t)}}\exp \left[ -\frac{z^2}
{2w_z^2(t)}+i\beta(t) z^2  \right],
\end{align}
where $w_z(t)$ is the width and $\beta(t)$ is the chirp. The  Lagrangian variational formalism leads to the following equation for the width $w_z(t)$ \cite{laserphysics}:
\begin{align}\label{vw1d}
\ddot{w}_{z}(t)
 +\lambda^2 w_z(t) =
\frac{1}{w_z^3(t)}+ \frac{ 2N_{\mathrm{at}}}{\sqrt{2\pi}}
\frac{ \left[{a}-a_{\mathrm{dd}}
c(\hat \kappa)\right]  }{d_\rho^2w_z^2(t)} , \quad \hat \kappa=\frac{d_\rho}{w_z(t)}.
\end{align}
The time-independent width of a stationary state can be obtained from Eq. (\ref{vw1d})
by setting $\ddot{w}_{z}(t)=0$.
The variational chemical potential for the stationary state  is given by \cite{laserphysics}
\begin{align}\label{vmu1d}
\mu = \frac{1}{4w_z^2}+\frac{2N_{\mathrm{at}}[a-a_{\mathrm{dd}}f(\hat \kappa)]}{\sqrt{2\pi}w_z d_\rho^2}
+\frac{\lambda^2 w_z^2}{4}.
\end{align}
The energy per particle is given by 
\begin{align}\label{ve1d}
\frac{E}{N_{\mathrm{at}}} = \frac{1}{4w_z^2}+\frac{N_{\mathrm{at}}[a-a_{\mathrm{dd}}f(\hat \kappa)]}{\sqrt{2\pi}w_z d_\rho^2}
+\frac{\lambda^2 w_z^2}{4}.
\end{align}

\subsubsection{$x$ direction}

For a cigar-shaped dipolar BEC with a strong axially-symmetric ($\nu=\lambda$)
radial
trap ($\gamma <\nu$, $\lambda$), 
we assume that the dynamics of the BEC 
in the radial direction is confined in the radial ground state \cite{laserphysics,santos1d,1d-dell}
\begin{align}
\phi(\boldsymbol \rho) = \exp(-\rho^2/2d_\rho^2)/
(d_\rho\sqrt\pi), \quad \nu d_\rho^2
=1,\quad \boldsymbol \rho \equiv (y,z),  
\end{align}
of the transverse trap and the wave 
function $\phi({\bf r})$ can be written as 
\begin{align}\label{an1x}
\phi({\bf r},t) \equiv 
\phi_{1D}(x,t) \times \phi(\boldsymbol \rho) =\frac{1}{\sqrt{\pi d_\rho^2}}\exp\left[  -\frac
{\rho^2}{2d_\rho^2}\right]  \phi_{1D}(x,t),
\end{align}
where $\phi_{1D}(x,t)$ is the effective 1D wave function for the dynamics along $x$ axis and 
$d_\rho$ is the radial harmonic oscillator length.

To derive the
effective 1D equation for the cigar-shaped dipolar BEC,
we substitute the ansatz (\ref{an1x}) in Eq. (\ref{gpe3d}), multiply by the
ground-state wave function $\phi(\boldsymbol \rho)$ and integrate over $\boldsymbol \rho$ to get
the 1D equation  
\begin{align}  \label{gpe1dx}
&
i\frac{\partial \phi_{1D}(x,t)}{\partial t}=\left[-\frac{\partial_x^2}{2}
+\frac{\gamma^2 x^2}{2}+\frac{2 aN_{\mathrm{at}} \vert  \phi_{1D} \vert  ^2}{ d_\rho^2}
+{4\pi a_{\mathrm{dd}}N_{\mathrm{at}}}\int_{-\infty}^{\infty}\frac{dk_x}{2\pi}e^{-ik_x x}\widetilde 
n (k_x,t)j_{1D}(\tau_x)\right] \phi_{1D}(x,t), 
\end{align}
where  $ \tau_x=d_\rho k_x/\sqrt 2$ and 
\begin{align}&
j_{1D}(\tau_x) \equiv\frac{1}{(2\pi)^2} \int d{\bf k}_\rho   \left[ \frac{3k_z^2}{{\bf k}^2} -1          \right]  | \widetilde n({\bf k}_\rho)|^2       
=\frac{\sqrt 2}{2 \pi d_\rho  }\int_{-\infty}^\infty d\tau_y e^{-\tau_y ^2}
h_{2D}(\tau)   ,  \quad \tau_y=\frac{d_\rho
k_y}{\sqrt 2}, 
\quad \tau=\sqrt{ \tau_x^2+\tau_y^2},\\
& h_{2D}(\tau)= \frac{1}{\sqrt{2\pi}d_{\rho}}[2-3\sqrt \pi e^{\tau^2}  \tau \{1- \text{erf}(\tau)\} ].
\end{align}
 To derive Eq. (\ref{gpe1dx}), the dipolar term in Eq. (\ref{gpe3d}) is first written in momentum space 
using Eq. (\ref{xyft}) and the integrations over $k_y$ and $k_z$ are  performed in the dipolar term.

\subsection{Two-dimensional GP equation for a disk-shaped dipolar BEC}

\label{2dgpe}
\subsubsection{$x-y$ plane} 
\label{2dgpexy}

For an axially-symmetric ($\nu=\gamma$) disk-shaped dipolar BEC with a strong axial trap 
 ($\lambda > \nu$, $\gamma$), 
we assume that the dynamics of the BEC in the axial direction
 is confined in the axial ground state
\begin{align}
\phi(z)=\exp(-z^2 /2d_z^2)/(\pi d_z^2)^{1/4}, \quad d_z= \sqrt{1/
(\lambda)},
\end{align}
and we have for the wave function
\begin{align} \label{an2}
\phi({\bf r})\equiv \phi(z) \times \phi_{2D}(\boldsymbol \rho,t)
=\frac{1}{(\pi d_z^2)^{1/4}}\exp\left[-\frac{z^2}{2d_z^2}  
\right] \phi_{2D}(\boldsymbol \rho,t),
\end{align}
where  $\boldsymbol \rho \equiv (x,y)  $,
$\phi_{2D}(\boldsymbol \rho,t)$ is the effective 2D wave function for the radial dynamics
and $d_z$ is the axial  harmonic oscillator length.
To derive the effective 2D equation for the disk-shaped dipolar BEC,
we use  ansatz (\ref{an2}) in Eq. (\ref{gpe3d}), multiply by the 
ground-state wave function $\phi(z)$ and integrate over $z$ to get 
the 2D equation \cite{laserphysics,santos2d}
 \begin{align}\label{gpe2d}
&
i\frac{\partial \phi_{2D}(\boldsymbol \rho,t)}{\partial t}=\left[-\frac{\nabla_\rho^2}{2}
+\frac{\gamma^2 x^2+\nu^2 y^2}{2}+\frac{4\pi aN_{\mathrm{at}} \vert  \phi_{2D} \vert  ^2}{\sqrt{2\pi}d_z}
+ { 4\pi }a_{\mathrm{dd}}N_{\mathrm{at}}\int \frac{d{\bf k}_\rho}{(2\pi)^2}
 e^{-i{\bf k}_\rho.\boldsymbol \rho}\widetilde n({\bf k}_\rho,t)h_{2D}\biggr(\frac{k_\rho d_z}{\sqrt 2}
\biggr)\right] \phi_{2D}(\boldsymbol \rho,t),
\end{align}
where $k_\rho=\sqrt{k_x^2+k_y^2}$, and  
\begin{align}
&\widetilde n({\bf k}_\rho,t)=\int d\boldsymbol \rho e^{i{\bf k}_\rho.\boldsymbol \rho} \vert  \phi_{2D}(\boldsymbol \rho,t) \vert  ^2,  \quad \widetilde n(k_z)=\int dz e^{i k_z z}|\phi(z)|^2= 
e^{-k_z^2d_z^2/4,}
\\
&
h_{2D}(\xi)\equiv \frac{1}{2\pi}\int^\infty_{-\infty} dk_z \left[  \frac{3k_z^2}{{\bf k}^2}  -1 \right] | \widetilde n(k_z) |^2=\frac{1}{\sqrt{2\pi}d_z} [2-3\sqrt \pi \xi \exp(\xi^2)\{1-\text{erf}(\xi)\}],\quad
 \xi=\frac{k_\rho d_z}{\sqrt 2}.
\end{align}
  To derive Eq. (\ref{gpe2d}), the dipolar term in Eq. (\ref{gpe3d}) is first written in momentum space 
using Eq. (\ref{xyft}) and the integration over $k_z$ is performed in the dipolar term.

The 2D GP equation (\ref{gpe2d}) can be solved analytically using the 
 Lagrangian variational formalism with the following Gaussian
ansatz for the 
wave function \cite{laserphysics}:
\begin{align}
 \phi_{2D}(\boldsymbol \rho,t)=\frac{\pi^{-1/2}}{w_\rho(t)}\exp \left[ -\frac{\rho^2}
{2w_\rho^2(t)}+i\alpha(t) \rho^2  \right],
\end{align}
where $w_\rho(t)$ is the width and $\alpha(t)$ is the chirp. The  Lagrangian variational formalism leads to the following equation for the width $w_\rho$ \cite{laserphysics}:
\begin{align}\label{vw2d}
\ddot{w}_{\rho}(t)
 +\gamma^2 w_\rho(t) =
\frac{1}{w_\rho^3(t)}+ \frac{ N_{\mathrm{at}}}{\sqrt{2\pi}}
\frac{ \left[2a-a_{\mathrm{dd}}
g(\bar \kappa)\right]  }{w_\rho^3(t)d_z} , \quad  \bar \kappa=\frac{w_\rho(t)}{d_z}.
\end{align}
The time-independent width of a stationary state can be obtained from Eq. (\ref{vw2d})
by setting $ \ddot{w}_{\rho}(t)=0$.
The variational chemical potential for the stationary state is given by \cite{laserphysics}
\begin{align}\label{vmu2d}
\mu = \frac{1}{2w_\rho^2}+\frac{2N_{\mathrm{at}}[a-a_{\mathrm{dd}}f(\bar \kappa)]}
{\sqrt{2\pi}d_z w_\rho^2}
+\frac{\gamma^2 w_\rho ^2}{2}.
\end{align}
The energy per particle is given by 
\begin{align}\label{ve2d}
\frac{E}{N_{\mathrm{at}}} = \frac{1}{2w_\rho^2}+\frac{N_{\mathrm{at}}[a-a_{\mathrm{dd}}f(\bar \kappa)]}
{\sqrt{2\pi}d_z w_\rho^2}
+\frac{\gamma^2 w_\rho ^2}{2}.
\end{align}

\subsubsection{$x-z$ plane} 
\label{2dgpexz}

For a  disk-shaped dipolar BEC with a strong axial trap along $y$ direction 
 ($\nu> \lambda$, $\gamma$), 
we assume that the dynamics of the BEC in the $y$ direction
 is confined in the ground state
\begin{align}
\phi(y)=\exp(-y^2 /2d_y^2)/(\pi d_y^2)^{1/4}, \quad d_y= \sqrt{1/
(\nu)},
\end{align}
and we have for the wave function
\begin{align} \label{an2xz}
\phi({\bf r})\equiv \phi(y) \times \phi_{2D}(\boldsymbol \rho{},t)
=\frac{1}{(\pi d_y^2)^{1/4}}\exp\left[-\frac{y^2}{2d_y^2}  
\right] \phi_{2D}(\boldsymbol \rho{},t),
\end{align}
where   now $\boldsymbol \rho{}\equiv (x,z)$, and 
$\phi_{2D}(\boldsymbol \rho{},t)$ is the circularly-asymmetric 
effective 2D wave function for the 2D dynamics
and $d_y$ is the   harmonic oscillator length along $y$ direction.
To derive the effective 2D equation for the disk-shaped dipolar BEC,
we use  ansatz (\ref{an2xz}) in Eq. (\ref{gpe3d}), multiply by the 
ground-state wave function $\phi(y)$ and integrate over $y$ to get 
the 2D equation  
 \begin{align}\label{gpe2dxz}
&
i\frac{\partial \phi_{2D}(\boldsymbol \rho,t)}{\partial t}=\left[-\frac{\nabla_\rho^2}{2}
+\frac{\gamma^2 x^2+\lambda^2 z^2}{2}+\frac{4\pi aN_{\mathrm{at}} \vert  \phi_{2D} \vert  ^2}{\sqrt{2\pi}d_y}
+ {4\pi a_{\mathrm{dd}}N_{\mathrm{at}}}\int \frac{d{\bf k}_\rho}{(2\pi)^2}
 e^{-i{\bf k}_\rho.\boldsymbol \rho}\widetilde n({\bf k}_\rho,t)j_{2D}
\biggr(\frac{k_\rho d_y}{\sqrt 2}\biggr)
\right] \phi_{2D}(\boldsymbol \rho,t),
\end{align}
where $k_\rho=\sqrt{k_z^2+k_x^2}$, and  
\begin{align}
&
j_{2D}(\xi)\equiv \frac{1}{2\pi}\int^\infty_{-\infty} dk_y \left[  \frac{3k_z^2}{{\bf k}^2}  -1 \right] |\widetilde n(k_y)|^2 =\frac{1}{\sqrt{2\pi}d_y}
[ -1+3\sqrt \pi \frac{\xi_z^2 }{\xi} \exp(\xi^2)\{1-\text{erf}(\xi)\}],\quad
\xi = \frac{k_\rho d_y}{\sqrt 2}, \xi_z=\frac{k_z d_y}{\sqrt 2}.
\end{align}
  To derive Eq. (\ref{gpe2dxz}), the dipolar term in Eq. (\ref{gpe3d}) is first written in momentum space 
using Eq. (\ref{xyft}) and the integration over $k_y$ is performed in the dipolar term.

\section{Details about the programs}

\label{programs}

\subsection{Description of the programs}

In this subsection we describe the numerical  codes for solving the dipolar 
GP equations (\ref{gpe1d}) and (\ref{gpe1dx}) in 1D, Eq.  (\ref{gpe2d}) and (\ref{gpe2dxz}) in 2D, and Eq. (\ref{gpe3d}) in 3D using real- and imaginary-time 
propagations. The real-time  propagation yields the time-dependent dynamical results  and the imaginary-time propagation yields the time-independent stationary
solution for the lowest-energy state for a specific symmetry.
We use the split-step Crank-Nicolson method for the solution 
of the equations described in Ref. \cite{CPC1}. The present programs have 
the same structure as in Ref. \cite{CPC1} with added subroutines to calculate 
the dipolar integrals. 
In the absence of dipolar interaction the present programs will be identical with the previously published ones \cite{CPC1}. 
A general instruction to use these programs in the nondipolar case   
can be found in Ref. \cite{CPC1} and we refer the interested reader to this article for the same. 

The present Fortran programs named 
(`imag1dX.f90', `imag1dZ.f90'),
(`imag2dXY.f90', `imag2dXZ.f90'),
`imag3d.f90', 
(`real1dX.f90', `real1dZ.f90'),
(`real2dXY.f90', `real2dXZ.f90'), 
`real3d.f90',
deal with imaginary- and real-time propagations in 1D, 2D, and 3D and   
are to be contrasted with previously published programs \cite{CPC1}
`imagtime1d.F',
`imagtime2d.f90',
`imagtime3d.f90',
`realtime1d.F',
`realtime2d.f90', and 
`realtime3d.f90', for the nondipolar case.
The input parameters in Fortran programs are introduced in the beginning of each program.
The corresponding C codes are called (imag1dX.c, imag1dX.h, imag1dZ.c, imag1dZ.h,), (imag2dXY.c, imag2dXY.h, imag2dXZ.c, imag2dXZ.h,), (imag3d.c, imag3d.h), (real1dX.c, real1dX.h, real1dZ.c, real1dZ.h,), (real2dXY.c, real2dXY.h, real2dXZ.c, real2dXZ.h,), (real3d.c, real3d.h), with respective 
input files  (`imag1dX-input', `imag1dZ-input'),
(`imag2dXY-input', `imag2dXZ-input'),
`imag3d-input', 
(`real1dX-input',`real1dZ-input'),
(`real2dXY-input', `real1dXZ-input'),
`real3d-input',
which perform identical executions as in the Fortran programs. 


We present in the following a description of input parameters. 
The  parameters NX, NY, and NZ in 3D (NX and NY in 2DXY, NX and NZ in 2DXZ),  and N in 1D 
stand for total number of space points in $x$, $y$ and
$z$ directions, where the respective space steps 
DX, DY, and DZ   can be made 
equal or different; DT is the time step. The parameters NSTP, NPAS, and NRUN
denote number of time iterations. 
The parameters GAMMA ($\gamma$), NU ($\nu$), and LAMBDA ($\lambda$) 
denote the anisotropy of the trap.  The number of atoms is denoted NATOMS ($N_{\mathrm{at}}$), 
the scattering length is denoted AS ($a$) and dipolar length ADD ($a_{\mathrm{dd}}$). 
The parameters G0 ($4\pi N_{\mathrm{at}} a$) and GDD0 ($3 a_{\mathrm{dd}} N_{\mathrm{at}} $) are the contact and dipolar 
nonlinearities.  
 The parameter OPTION = 2 (default) defines the equations of the present paper with a factor 
of half before the kinetic energy and trap; OPTION = 1 defines a different 
set of GP equations without these factors, viz Ref. \cite{CPC1}.
 The parameter AHO is the unit of length 
and Bohr\_a0 is the  Bohr radius. 
In 1D the parameter DRHO is the radial harmonic oscillator 
 $d_\rho$ and in 2D the parameter 
D$\_$Z or D$\_$Y is the axial harmonic oscillator length $d_z$ or $d_y$. 
The parameter CUTOFF is the cut-off $R$ of Eq. (\ref{cutoff}) in the 3D programs. The 
parameters GPAR and GDPAR are constants which multiply the nonlinearities G0 and GDD0 
in realtime routines before NRUN time iterations to study the dynamics.

The programs, as supplied, solve the GP equations for  specific values of
dipolar and contact  nonlinearities and write the wave function, chemical potential,
energy, and root-mean-square (rms) size(s) etc.
For solving a stationary  problem, the imaginary-time programs are far more accurate and should be used. The real-time
programs should be used for studying non-equilibrium problems reading an initial wave function calculated by the imaginary-time
program with identical set of parameters (set NSTP = 0, for this purpose, in the real-time programs). The real-time programs can also calculate stationary solutions in NSTP time steps (set NSTP $\ne$ 0 in real-time programs), however, with 
less accuracy compared to the imaginary-time programs. The larger the value of NSTP in real-time programs, more accurate will be the result \cite{CPC1}. 
The nonzero integer parameter NSTP refers to the number of time iterations during which the nonlinear terms are slowly introduced 
during the
time propagation for calculating the wave function.
After introducing the nonlinearities in NSTP iterations
the imaginary-time programs calculate the final result in NPAS plus NRUN time steps and write some of the results after NPAS steps to check convergence. The real-time programs run the dynamics during NPAS steps with unchanged initial parameters so as to check the stability and accuracy of the results. Some of the nonlinearities are then slightly modified after NPAS iterations and the small oscillation of the system is studied during NRUN iterations.

Each program is preset at  fixed values of  contact and dipolar
nonlinearities as calculated from input scattering length(s), dipolar strength(s),
and number of atom(s),  correlated DX-DT values and NSTP, NPAS, and NRUN, etc.
A study of the correlated DX and DT values in the nondipolar case can be found in Ref. \cite{CPC1}.
 Smaller the steps DX, DY, DZ
and DT, more accurate will be the result,  provided we integrate 
over a reasonably  large space region by increasing NX, NY, and NZ, etc.  
Each supplied program produces result up to a desired precision consistent with the parameters employed $-$ G0, GDD0, DX, DY, DZ, DT, NX,
NY, NZ,
NSTP, NPAS, and NRUN  etc.

\subsection{Description of Output files}

Programs `imag$n$d*' ($n = 1,2,3,$ C and Fortran):  They write  final density in files 
`imag$n$d-den.txt'  after NRUN iterations. In addition,  in 2D and 3D, integrated 
1D densities  `imag$n$d*-den1d\_x.txt', `imag$n$d*-den1d\_y.txt',`imag$n$d*-den1d\_z.txt', along 
x, y, and z, etc,  are given. These densities are obtained by integrating the 
densities over eliminated space variables. 
In addition, in 3D integrated 2D densities `imag3d-den2d\_xy.txt', `imag3d-den2d\_yz.txt',`imag3d-den2d\_zx.txt', in xy, yz, and zx planes can be written (commented out by default).
The files `imag$n$d*-out.txt'  provide different initial input data,   as well as chemical potential, energy, size, etc at different stages (initial, after NSTP, NPAS, and after NRUN iterations), from which a convergence of the result can be inferred.  The files 
`imag$n$d*-rms.txt' provide the different rms sizes at different stages (initial, after NSTP, NPAS, and after NRUN iterations).

Programs `real$n$d*'  ($n = 1,2,3,$ C and Fortran):  The same output files as in the 
imaginary-time programs are available in the real-time programs. The real-time densities are 
reported after NPAS iterations.   In addition 
in the `real$n$d*-dyna.txt'  file the temporal evolution of the widths are given during NPAS and NRUN iterations. Before NRUN iterations the nonlinearities G0 and GDD0 are multiplied by parameters GPAR and GDPAR to start an oscillation dynamics.

\subsection{Running the programs}

In addition to installing the respective Fortran  and C compilers  one needs 
also to install the FFT routine FFTW in the computer. 
To run the  Fortran programs the supplied routine fftw3.f03 should be included in 
compilation. The commands 
for running the Fortran programs using INTEL, GFortran, and Oracle Sun compilers 
are given inside the Fortran programs. 
The programs are submitted in directories with option to compile using the command `make'.
There are two files with general information about the programs and FT
for user named 'readme.txt' and `readme-fftw.txt'. The Fortran and 
C programs are in directories ../f$\_$program and   ../c$\_$program.  Inside these directories 
there are subdirectories such as ../input, ../output,  ../src. The subdirectory  ../output contains output files the programs generate, ../input contains input files for C programs, and 
../src contains the different programs. The command `make' in the directory ../f$\_$program or 
 ../c$\_$program compiles all the programs and generates the corresponding executable files to run. 
The command `make' for INTEL, GFortran and OracleSun Fortran  are  given.

\section{Numerical Results}

\label{num}

In this section we present results for energy, chemical potential and root-mean-square (rms)
sizes for 
different stationary BECs in 1D, 2D, and 3D, and compare 
with those obtained by using Gaussian variational and TF approximations, wherever possible. 
We also compare with available results by other authors.    
For a fixed space and time step, sufficient number of space discretizing points and 
time iterations are to be allowed to get convergence.  

\begin{table}[!ht]
\caption{The energy per particle  $E/N_{\mathrm{at}}$, chemical potential $\mu$, and rms size $\langle
z \rangle$ of the 1D GP equation (\ref{gpe1d})  
for $\lambda=1, d_\rho = 1$ $\mu$m for the $^{52}$Cr BEC with $a=6 $ nm, $a_{\mathrm{dd}}
=16a_0$ and different number of atoms $N_{\mathrm{at}}$. In Eqs. (\ref{gpe3d}) and 
(\ref{gpe1d})
the 
lengths are expressed in oscillator unit:     $l=1$ $\mu$m. 
Numerical results are calculated for parameters
(A) $dz=0.05 ,dt=0.0005, N=2048$ 
(B) $dz=0.1 , dt=0.001, N=1024$   
  and compared with 
variational results obtained from Eqs. (\ref{vw1d}), (\ref{vmu1d}) and (\ref{ve1d}).  }
\label{table1}
\begin{center}
\begin{tabular}{rrrrrrrrrr}
\hline\hline
\multicolumn{1}{c}  {$N_{\mathrm{at}}$}
        & \multicolumn{1}{c}{$\langle z\rangle $}  
        & \multicolumn{1}{c}{$\langle z\rangle$}  
        & \multicolumn{1}{c}{$\langle z\rangle$}  
     & \multicolumn{1}{c}{$E/N_{\mathrm{at}}$}
        & \multicolumn{1}{c}{$E/N_{\mathrm{at}}$}
        & \multicolumn{1}{c}{$E/N_{\mathrm{at}}$}
        & \multicolumn{1}{c}{$\mu$} 
        & \multicolumn{1}{c}{$\mu$} 
        & \multicolumn{1}{c}{$\mu$} 
\\
  &var   &(B)  
  & (A)  &var & (B)&(A)   
&    var  &(B)&(A) 
\\
\hline
100 &  0.7939&     0.7937 &0.7937  &0.7239   & 0.7222 &0.7222   &  0.9344 &
0.9297&0.9297
\\
500 &1.0425&   1.0381  &1.0381  &1.4371  &1.4166  &1.4166  &2.2157  
&2.1691 &2.1691
\\
1000 &1.2477  &  1.2375 &1.2375  & 2.1376
  &  2.0920  & 2.0920&3.4165
&3.3234&3.3234
\\
5000 &2.0249      & 1.9939  &1.9939 &5.8739   &  5.6910&5.6910  &9.6671  
& 9.3488& 9.3488
\\
10000 & 2.5233    & 2.4815  & 2.4815 &9.2129    &8.913 &8.913
& 15.223 
&  14.715&14.715
\\
50000&  4.2451     &   4.1719&4.1719 &  26.505 &25.622  &  25.622  & 43.993 
&42.527 &42.527
\\
\hline\hline
\end{tabular}
\end{center}

\end{table}

First we present in table \ref{table1}
numerical results for the energy $E$, chemical potential $\mu$, 
and rms size $\langle z \rangle$ calculated using
the imaginary-time program for  
the 1D dipolar GP equation (\ref{gpe1d})
for $^{52}$Cr atoms with $a=6$ nm ($\approx 113a_0$ with $a_0$ the Bohr radius), and $a_{\mathrm{dd}}=16a_0$ for $\lambda =1, d_\rho =1, l =1$
 $\mu$m and for different number of atoms $N_{\mathrm{at}}$ and different space and time steps 
$dz$ and $dt$. 
The Gaussian variational approximations obtained from Eqs. (\ref{vw1d}), (\ref{vmu1d}) and (\ref{ve1d})
are also given for comparison. The  
variational results provide better approximation to the numerical solution 
for a smaller number of atoms. 

\begin{table}[!ht]
\caption{The energy per particle $E/N_{\mathrm{at}}$, chemical potential $\mu$, and rms size $\langle
\rho \rangle$ of the 2D GP equation (\ref{gpe2d})  
for $\gamma=\nu=1, d_z = 1$ $\mu$m for the $^{52}$Cr BEC with $a=6 $ nm, $a_{\mathrm{dd}}
=16a_0$ and different number of atoms $N_{\mathrm{at}}$. In Eqs. (\ref{gpe3d}) and 
(\ref{gpe2d})
the 
lengths are expressed in oscillator unit:     $l=1$ $\mu$m. 
Numerical results are calculated for space and time steps
(A) $dx=dy=0.1 , dt=0.0005,  NX=NY\equiv {\cal N} =768,  $  
(B) $dx=dy=0.2, dt=0.002, {\cal N}=384,$ and compared with 
variational results obtained from Eqs. (\ref{vw2d}), (\ref{vmu2d}) and (\ref{ve2d}).}
\label{table2}
\begin{center}
\begin{tabular}{rrrrrrrrrrr}
\hline\hline
\multicolumn{1}{c}  {$N_{\mathrm{at}}$}
        & \multicolumn{1}{c}{$\langle\rho\rangle $}  
        & \multicolumn{1}{c}{$\langle\rho\rangle$}  
        & \multicolumn{1}{c}{$\langle\rho\rangle$}   
     & \multicolumn{1}{c}{$E/N_{\mathrm{at}}$}
        & \multicolumn{1}{c}{$E/N_{\mathrm{at}}$}
        & \multicolumn{1}{c}{$E/N_{\mathrm{at}}$}
        & \multicolumn{1}{c}{$\mu$} 
        & \multicolumn{1}{c}{$\mu$} 
        & \multicolumn{1}{c}{$\mu$} 
\\
  &var  & (B) &(A)  
  & var  & (B)& (A)   
&    var  & (B)&(A) 
\\
\hline
100 &1.0985   & 1.097&1.097   &1.2182 &1.2156  & 1.2157 & 1.4187  & 
1.4120&1.4119
\\
500 &1.3514   & 1.342 &1.342  & 1.8653& 1.8383 &  1.8383  &  2.5437   & 2.4840&
2.4840
\\
1000 &1.5482 & 1.530  &  1.531 &  2.4571 &2.3988  &2.3988  & 3.5070& 
3.3901&3.3901
\\
5000 &2.2549   &2.208  & 2.208   & 5.2206 & 4.9989 &  4.9989& 7.8005  & 7.4249 &7.4249
\\
10000 &2.6824   &2.619& 2.619   &7.3787 &7.029  & 7.029   & 11.090 & 10.522 & 
10.522
\\
50000& 4.0420   &3.934& 3.934   &  16.680 &15.793  & 15.793&  25.161  &
  23.789 &23.789
\\
\hline\hline
\end{tabular}
\end{center}

\end{table}

In table \ref{table2} we present results for the energy $E$, chemical potential $\mu$, 
and rms size $\langle \rho \rangle$ of the 2D GP equation (\ref{gpe2d}) for $\gamma =\nu = 1, d_z =1,  l=1$ 
$\mu$m. The numerical results are calculated using different space and time steps $dx, dy$ 
and $dt$ and different number $N_{\mathrm{at}}$ of $^{52}$Cr atoms with $a_{\mathrm{dd}}=16a_0$ and $a=6$ nm. 
Axially-symmetric Gaussian variational approximations obtained from   Eqs. (\ref{vw2d}), (\ref{vmu2d}) and (\ref{ve2d})
are also presented for comparison.

\begin{table}[!ht]
\caption{Energy per particle $E/N_{\mathrm{at}}$ and chemical potential $\mu$ from a solution of Eq. 
(\ref{gpe3d}) for $\gamma=\nu=1, \lambda^2 = 0.25, a=0$ 
and different nonlinearity $g_{\mathrm{dd}}\equiv 3 a_{\mathrm{dd}}N_{\mathrm{at}}$.  The present numerical 
results are compared with 
Gaussian variational 
results obtained from  Eqs.   (\ref{ve3d}) and (\ref{vmu3d})  
as well as numerical results  of Asad-uz-Zaman {\it et al.} \cite{Blume,Blume2}. 
Numerical results are calculated for the following space and time steps 
and the following space discretizing points in the Crank-Nicolson
discretization:
(A) $dx=dy=dz=0.05 , dt=0.0004,$ $(NX=NY=NZ\equiv {\cal N}=384) $; 
(B) $0.1 , dt=0.002,$ $ ({\cal N}=128, R =6) $; 
 and (C) $0.2 ,dt=0.007,$ $ ({\cal N}=64, R=6)$. }
\label{table3}
\begin{center}
\begin{tabular}{rrrrrrrrrrrrr}
\hline\hline
\multicolumn{1}{c}  {$ g_{\mathrm{dd}}$}
        & \multicolumn{1}{c}{$E/N_{\mathrm{at}}$}  
        & \multicolumn{1}{c}{$E/N_{\mathrm{at}}$}  
        & \multicolumn{1}{c}{$E/N_{\mathrm{at}}$}  
        & \multicolumn{1}{c}{$E/N_{\mathrm{at}}$ } 
   & \multicolumn{1}{c}{$ E/N_{\mathrm{at}}$}  
        & \multicolumn{1}{c}{$ \mu$}
        & \multicolumn{1}{c}{$ \mu$} 
        & \multicolumn{1}{c}{$ \mu$} 
        & \multicolumn{1}{c}{$ \mu$} 
   & \multicolumn{1}{c}{$ \mu$} 
\\
    & var & (C)  &  (B)& (A)  & \cite{Blume2} & var   & (C)  & (B)
& (A) 
&\cite{Blume2}\\
\hline
0    & 1.2500&1.2498   &1.2500&1.2500  &
   1.2500     & 1.2500   & 1.2498  &1.2500&1.2500   &1.2500   \\
1    & 1.2230&{1.2220}   & {1.2222} 
&1.2222 &1.2222     & 1.1934   &{1.1910}  
&  {1.1912}&1.1911  &1.1911   \\
2   & 1.1907&{1.1872}   &{1.1875}& 1.1874    &1.1874  &  1.1203     & {1.1100}
& {1.1100}& 1.1100  &1.1100  \\
3    &  1.1521  & {1.143}   &
{1.1439}  &  1.1438  & 1.1437  & 1.0253    & {0.995}  
&{0.996}& 0.996  & 0.9955   \\
4    &  1.1051 & { 1.085}  & {1.0857} & 1.0857 &  1.0856   &   0.8950     &  {0.805}   
&{0.803} &0.806
&  0.8062   \\
\hline\hline
\end{tabular}
\end{center}

\end{table}

Now we present results of the solution of the 3D GP equation (\ref{gpe3d}) with some axially-symmetric traps.  
In this case we take advantage of the cut-off introduced in Eq. (\ref{cutoff}) to improve 
the accuracy of the numerical calculation. The cut-off parameter $R$ was taken larger than the 
condensate size and smaller than the discretization box. 
First we consider the model 
3D GP equation with $a=0$ and different $g_{\mathrm{dd}}=3 a_{\mathrm{dd}}N_{\mathrm{at}}=1,2,3,4$ in an axially-symmetric trap 
with $\lambda =1/2$ and $ \nu=\gamma=1$.  The numerical results 
for different 
number of space  and time steps together with Gaussian variational results obtained from Eqs.   (\ref{ve3d}) and (\ref{vmu3d})
are 
shown in table \ref{table3}. These results for energy $E$ and chemical potential $\mu$ are compared with 
those calculated by Asad-uz-Zaman {\it et al.} \cite{Blume,Blume2}.  The present calculation is performed in the Cartesian $x,y,z$ coordinates and the dipolar term is evaluated by FT
 to momentum space. Asad-uz-Zaman {\it et al.} take advantage of the axial symmetry and perform the calculation in the axial $\rho,z$ ($\rho \equiv {x,y}$) variables 
and evaluate the dipolar term by a combined Hankel-Fourier transformation to momentum space for $\rho$ and $z$, respectively. The calculations of  Asad-uz-Zaman {\it et al.}
for stationary states involving two variables ($\rho$ and $z$)  thus could be more economic and accurate than the present calculation involving three Cartesian variables for the axially-symmetric configuration considered in table {\ref{table3}}.
 However, the present method, unlike that of Ref. \cite{Blume},   is readily applicable to the 
fully asymmetric configurations. Moreover, 
the present calculation for dynamics (non-stationary states) in 3D are more realistic  than the calculations of Asad-uz-Zaman {\it et al.}, where one degree of freedom is frozen.  For edxample, 
a votex could be unstable  \cite{afta} in a full 3D calculation, whereas a 2D calculation could make the same vortex stable.

\begin{table}[!ht]
\caption{ Energy per particle $E/N_{\mathrm{at}}$, and chemical potential $\mu$ from a solution of Eq. 
(\ref{gpe3d}) for $\gamma=\nu=1, \lambda^2 = 0.25, 4\pi a=0.20716, 4\pi a_{\mathrm{dd}}
= 0.033146$ and different number $N_{\mathrm{at}}$ of atoms. These nonlinearity parameters
taken from Ref. \cite{Bao2010} 
correspond to a $^{52}$Cr dipolar BEC with   $a\approx 100 a_0$ and 
$a_{\mathrm{dd}}\approx 16 a_0$ and oscillator length $l\approx 0.321$ $\mu$m. 
Variational and  TF results 
as well as numerical results  of Bao {\it et al.} \cite{Bao2010} 
are also shown. 
Numerical results are calculated using the following space and time steps 
and the following space discretizing points in the Crank-Nicolson
discretization:
(A) $dx=dy=dz=0.15 ,dt=0.002;$  
(B) $dx=dy=dz=0.3, dt=0.005. $ 
In (A) we take $NX=NY=NZ\equiv {\cal N}=128, R=9$ for $N_{\mathrm{at}}=100,500, 1000$
and ${\cal N}= 192, R=14$, for $N_{\mathrm{at}}=5000,10000,  50000$ ; and in (B) we take ${\cal N}= 64, R=9$, for $N_{\mathrm{at}}=100,500, 1000$
and ${\cal N}= 96, R=14$, for $N_{\mathrm{at}}=5000,10000,  50000$.
}
\label{table4}
\begin{center}
{ 
\begin{tabular}{rrrrrrrrrrrr}
\hline\hline
\multicolumn{1}{c}  {$N_{\mathrm{at}}$}
        & \multicolumn{1}{c}{$E/N_{\mathrm{at}}$}  
        & \multicolumn{1}{c}{$E/N_{\mathrm{at}}$}  
        & \multicolumn{1}{c}{$E/N_{\mathrm{at}}$}    
 & \multicolumn{1}{c}{$E/N_{\mathrm{at}}$ }   
& \multicolumn{1}{c}{$E/N_{\mathrm{at}}$ }
        & \multicolumn{1}{c}{$\mu$}
        & \multicolumn{1}{c}{$\mu$} 
        & \multicolumn{1}{c}{$\mu$} 
        & \multicolumn{1}{c}{$\mu$} 
 & \multicolumn{1}{c}{$\mu$} 
\\
    & var&TF & (B)  &  (A)  & \cite{Bao2010} & var & TF  & (B)  & (A) 
&\cite{Bao2010}\\
\hline
100    & 1.579& 0.945  &{1.567}   &{1.567}
  & 1.567     & 1.840   & 1.322&{1.813}  &{1.813}
& 1.813   \\
500    & 2.287&1.798  &{2.224} & {2.224}&  2.225     & 2.951   & 2.518&{2.835}  
& { 2.835} &2.837   \\
1000   & 2.836& 2.373 &{2.728}  &{2.728}  & 2.728    & 3.767     
&3.322& {3.583}  & {3.582} &3.583  \\
5000    &5.036& 4.517 & {4.744}   &{4.744} &  4.745  & 6.935      &6.324& {6.485}  
&6.486&    6.488   \\
10000    &6.563&5.960 &  {6.146}  &{6.146}  & 6.147  &9.100       &8.344&  {8.475} 
& {8.475}   &
  8.479   \\
50000    &12.34& 11.35&{11.46}   &{11.46} &   11.47  & 17.23      &15.89&
{15.96}
& {15.97}&   15.98   \\
\hline\hline
\end{tabular}
}
\end{center}

\end{table}

\begin{table}[!ht]
\caption{The rms sizes  $\langle x \rangle$ and $\langle z \rangle$ 
for the same systems illustrated in table  \ref{table4} using the same cut-off parameter $R$.}
\label{table5}
\begin{center}
{
\begin{tabular}{rrrrrrrrrrrr}
\hline\hline
\multicolumn{1}{c}  {$N$}
        & \multicolumn{1}{c}{$\langle z\rangle $}  
        & \multicolumn{1}{c}{$\langle z
\rangle$}  
        & \multicolumn{1}{c}{$\langle z
\rangle$}  
        & \multicolumn{1}{c}{$\langle z \rangle$}  
        & \multicolumn{1}{c}{$\langle z\rangle$}
        & \multicolumn{1}{c}{$\langle x\rangle$}
        & \multicolumn{1}{c}{$\langle x\rangle$}
        & \multicolumn{1}{c}{$\langle x\rangle$} 
        & \multicolumn{1}{c}{$\langle x\rangle$} 
        & \multicolumn{1}{c}{$\langle x\rangle$} 
\\
  & TF &var&(B)  & (A)   &  \cite{Bao2010}
  & TF & var&(B)  & (A)   &  \cite{Bao2010}
\\
\hline
100 &1.285   & 1.316&{1.305}   &{1.303}&  1.299 & 0.600   & 0.799
& {0.794}&{0.795} &0.796 
\\
500 &1.773   & 1.797&{1.752}  
&{ 1.752}& 1.745  &0.828   & 0.952   
&{0.938}  & {0.939}&    0.940
\\
1000 &2.037  & 2.079&  {2.014} &{2.014}   &2.009  & 0.951& 1.054     
&{1.035}  &{1.035}   & 1.035
\\
5000 &2.810   &2.904& {2.795}   &{2.795}&   2.790&1.313   & 1.392      
& {1.353}  &{1.353} &1.354
\\
10000 &3.228   &3.345& {3.217}   &{3.216}   &3.212& 1.508 &1.586&  {1.537}     
&{1.537}&   1.538  
\\
50000& 4.454   &4.629& {4.450}   &{4.450} & 4.441&2.080  &2.171 & {2.093} 
&{2.093}&   2.095 
\\
\hline\hline
\end{tabular}}
\end{center}

\end{table}

\begin{table}[!ht]
\caption{Energy per particle $E/N_{\mathrm{at}}$, chemical potential $\mu$, and rms sizes from a solution of Eq. 
(\ref{gpe3d}) for $^{52}$Cr atoms with  $\gamma=1, \nu^2=1/ 2, \lambda^2 = 1/4, a=110a_0$, 
$a_{\mathrm{dd}}=16a_0$, and harmonic oscillator length $l=1$ $\mu$m for different $N_{\mathrm{at}}$.  
Numerical results are calculated using the following space and time steps 
and the following space discretizing points in the Crank-Nicolson
discretization: 
(A) $dx=dy=dz=0.1 , dt=0.001$; 
 and (B) $0.2, dt=0.003 $.
In (A) we take $NX=NY=NZ\equiv {\cal N}=128, R=6$ for $N_{\mathrm{at}}=100,500, 1000$
and ${\cal N}=256 , R=10.5$, for $N_{\mathrm{at}}=5000,10000,  50000$ ; and in (B) we take ${\cal N}= 64, R=6$,
 for   $N_{\mathrm{at}}=100,500, 1000$
and ${\cal N}= 128, R=12$, for $N_{\mathrm{at}}=5000,10000,  50000$.
}
\label{table6}
\begin{center}
{
\begin{tabular}{rrrrrrrrrrr}
\hline\hline
\multicolumn{1}{c}  {$ N$}
        & \multicolumn{1}{c}{$E/N_{\mathrm{at}}$}  
        & \multicolumn{1}{c}{$E/N_{\mathrm{at}}$}  
        & \multicolumn{1}{c}{$\mu$ } 
        & \multicolumn{1}{c}{$ \mu$}
        & \multicolumn{1}{c}{$\langle x
\rangle $} 
        & \multicolumn{1}{c}{$\langle y
\rangle$} 
        & \multicolumn{1}{c}{$ \langle z
\rangle$} 
   & \multicolumn{1}{c}{$\langle x
\rangle$}
& \multicolumn{1}{c}{$\langle y
\rangle$}
& \multicolumn{1}{c}{$\langle z
\rangle$} 
\\
& (B) & (A) & (B) &(A) & (B) &(B) &(B) &(A) &(A) &(A) \\
\hline\hline
100 &1.219 &1.219 & 1.321 &1.321 & 0.742 &0.901 & 1.120& 0.742&0.901 &1.119 \\
500 &1.525 & 1.525  &1.830 & 1.830  & 0.818& 1.032 &1.379 &0.818&1.032 &1.379 \\
1000 &1.784 &1.784  &2.232 & 2.232& 0.874 &1.128 &1.559 &0.874 &1.129 & 1.558\\
5000 &2.885 &2.885  &3.857 &3.858 &  1.079 &1.463 & 2.132& 1.079&1.463 &2.132 \\
10000 &3.673 &3.673  &4.992 &4.992  &1.206&1.660 &2.450 & 1.206&1.660 &2.449 \\
50000 &6.713 &6.713 &9.306 &9.306   & 1.609 &2.260 &3.383 & 1.609&2.260 &3.383 \\
\hline
\end{tabular}}
\end{center}

\end{table}

Next we consider the solution of the 3D GP equation (\ref{gpe3d})
for a model condensate of $^{52}$Cr atoms in a
cigar-shaped  
axially-symmetric trap with $\gamma=\nu =1, \lambda = 1/2$, first considered by 
Bao {\it et al.} \cite{Bao2010}. The nonlinearities considered there 
 ($4\pi a = 0.20716, 4\pi a_{\mathrm{dd}} = 0.033146$) 
correspond to the following approximate values of $a, a_{\mathrm{dd}}$ and $l$:
$a\approx 100a_0, a_{\mathrm{dd}}
\approx 16a_0$, and $l=0.321$ $\mu$m. We present results for energy $E$ and chemical potential $\mu$ in table 
\ref{table4} and rms sizes $\langle z \rangle$ and $\langle x \rangle $ in table \ref{table5}. We also present 
variational and Thomas-Fermi (TF) results in this case together with results of numerical calculation of 
Bao {\it et al.} \cite{Bao2010}. The TF energy and chemical potential
in table \ref{table4} are calculated using Eqs. (\ref{tfen}) and (\ref{tfch}), respectively.
The TF sizes $\langle x\rangle $ and $\langle z\rangle $ in table \ref{table5} are obtained
 from Eqs. (\ref{ktf}) and (\ref{rtf}) using the TF density (\ref{tfden}).
  For small nonlinearities or small number of atoms, the Gaussian variational results obtained from Eqs. (\ref{f1}), (\ref{f2}),
 (\ref{ve3d}), and (\ref{vmu3d})
are in good agreement with the numerical calculations as the wave function for small  
nonlinearities has a quasi-Gaussian shape. However, for large nonlinearities or large number of atoms, the 
wave function has an approximate TF shape (\ref{tfden}), and the TF results provide better approximation 
to the numerical results, as can be seen from tables \ref{table4} and \ref{table5}.

After the consideration of 3D axially-symmetric trap now we consider a fully anisotropic trap in 3D. 
In table \ref{table6} we present the results for energy $E/N_{\mathrm{at}}$, chemical potential $\mu$ and rms sizes 
$\langle x \rangle,  \langle y \rangle,\langle z \rangle$ of 
a $^{52}$Cr BEC in a fully anisotropic trap with $\gamma=1, \nu =1/\sqrt 2, \lambda=1/2$ for different 
number of atoms. In this case we take $a=110a_0, a_{\mathrm{dd}}=16a_0$ and $l=1$ $\mu$m. The convergence of the calculation is studied by taking reduced space and time steps $dx$ and $dt$ and different number of 
space discretization points. Sufficient number of time iterations are to be allowed in each case to 
obtain convergence. In 3D the estimated numerical error in the calculation is less than 
0.05$\%$.
The error is associated with the intrinsic accuracy of the FFT routine for long-range dipolar 
interaction.

\begin{figure}[!t] 
\begin{center}  
\includegraphics[width=.495\linewidth,clip]{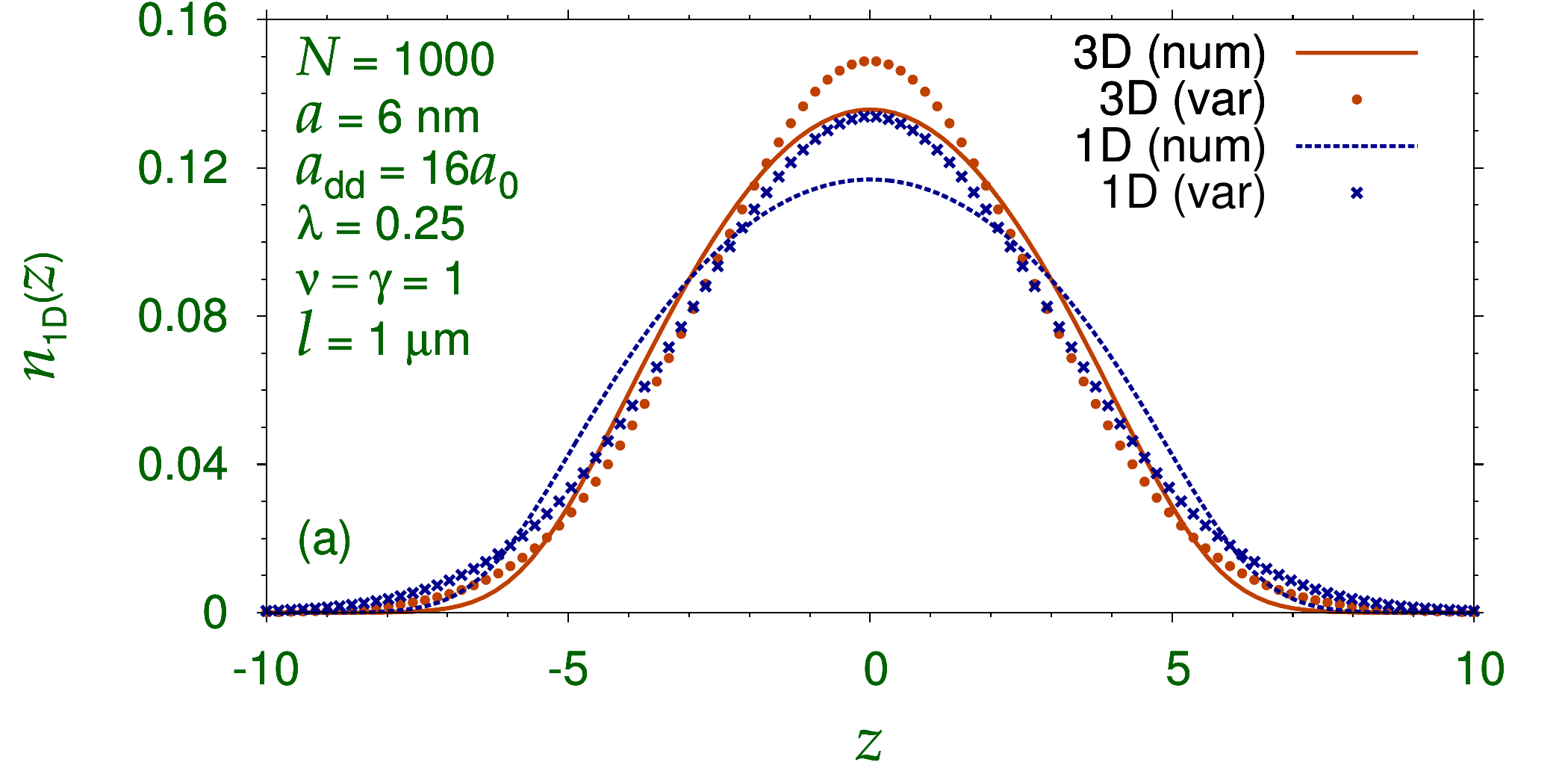} 
\includegraphics[width=.495\linewidth,clip]{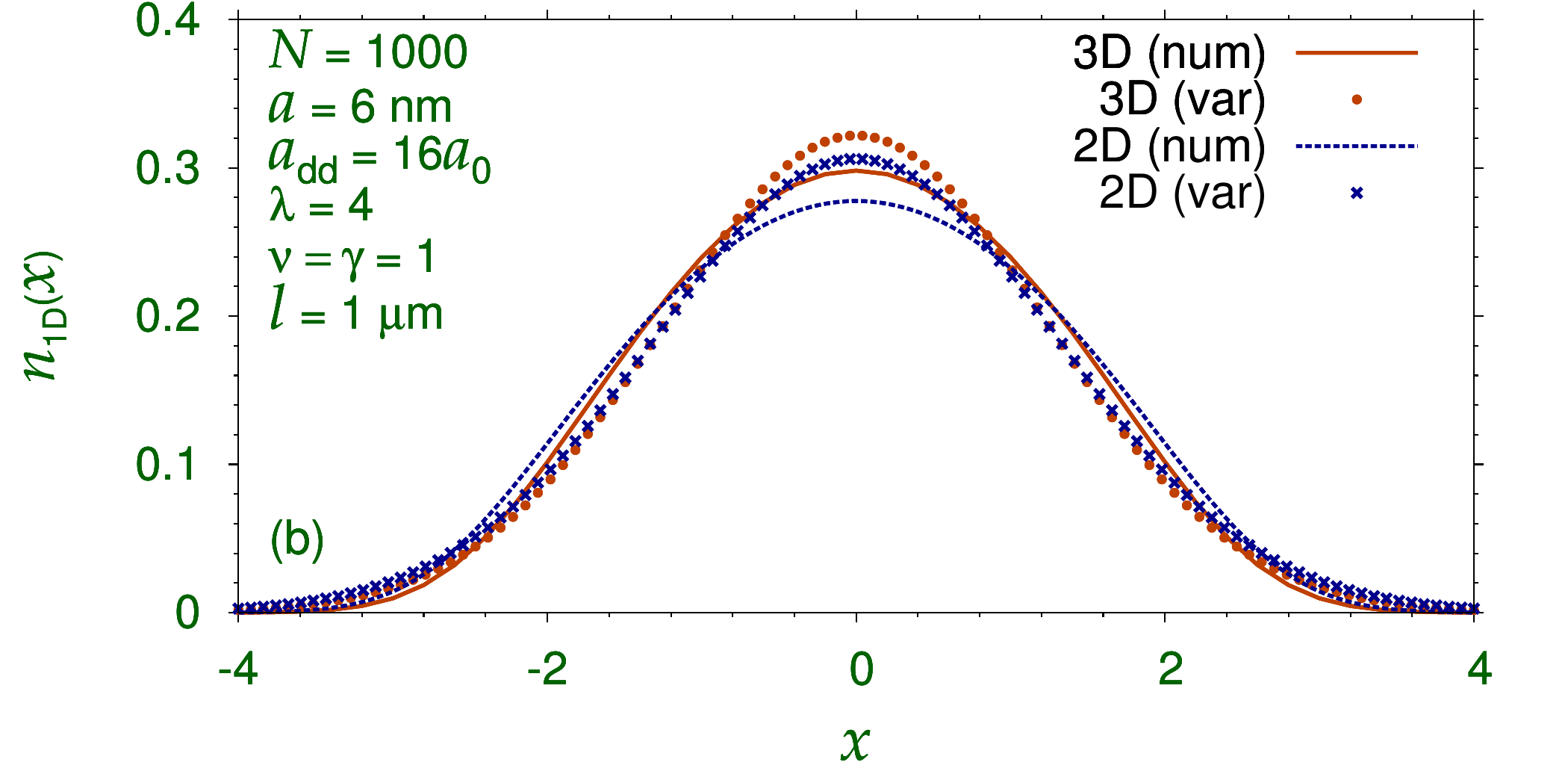} 
\end{center}
\caption{(a) Numerical (num) and variational (var) results for the
one-dimensional axial density $n_{1D}(z)= |\phi_{1D}(z)|^2$ along $z$ axis
for $\nu=\gamma=1, \lambda =0.25$
 of a cigar-shaped BEC of $N_{\mathrm{at}}=1000$ atoms obtained using the 
1D Eq. (\ref{gpe1d}) 
and that obtained after integrating the 3D density from Eq. (\ref{gpe3d}) over $x$ and $y$: 
$n_{1D}(z)= \int |\phi({\bf r})|^2 dx dy$.
(b) Numerical (num) and variational (var) results for the  1D radial density 
$n_{1D}(x) =\int |\phi({\bf r})|^2 dy dz
$ along $x$ axis
for $\nu=\gamma=1, \lambda =4$
 of a disk-shaped BEC of $N_{\mathrm{at}}=1000$ atoms obtained after integrating the 3D density from Eq. (\ref{gpe3d}) over $y$ and $z$ and after integrating the 2D density from Eq. (\ref{gpe2d}) over $y$ 
as follows:
 $n_{1D}(x)=\int dy |\phi_{2D}(x,y)|^2$
and $n_{1D}(x)=\int dy dz |\phi_{3D}(x,y,z)|^2$. In all cases $a=6$ nm and $a_{dd}= 16a_0$.
}
\label{fig1}
\end{figure}

The 1D and 2D GP equations (\ref{gpe1d}) and (\ref{gpe2d}) are valid for cigar- and disk-shaped BECs, respectively. In case of cigar shape the 1D GP  equation yields results for axial density and in this case it is appropriate to compare this density with the reduced axial density obtained by integrating the  3D density  over radial coordinates: 
$\displaystyle n(z)\equiv  \vert  \varphi (z) \vert  ^2= \int  \vert  \phi(x,y,z) \vert  ^2 \, dx\, dy $. In Fig. \ref{fig1} (a) we compare two axial densities
obtained from the 1D and 3D GP equations. We also show the densities calculated from 
the Gaussian variational approximation in both cases. 
   In the cigar case the trap parameters are $\nu  = \gamma
 = 1, \lambda  = 1/4$.
Similarly, for the disk shape it is interesting to compare the density along the radial 
direction in the plane of the disk as obtained from the 3D equation (\ref{gpe3d}) and the 2D equation (\ref{gpe2d}). 
In this case it is appropriate to calculate the 1D radial density along, say, $x$ direction by  
integrating 2D and 3D densities as follows: $n_{1D}(x)=\int dy |\phi_{2D}(x,y)|^2$
and $n_{1D}(x)=\int dy dz |\phi_{3D}(x,y,z)|^2$.   
In Fig. \ref{fig1} (b) we compare two radial densities
obtained from the 2D and 3D GP equations.  
We also show the densities calculated from 
the Gaussian variational approximation in both cases. 
 For this illustration, we consider the    trap parameters $\nu  = \gamma
 = 1, \lambda  = 4$. In both Figs. \ref{fig1} (a) and (b), the densities obtained 
from the solution of the 3D GP equation are in satisfactory  agreement with those obtained from a solution of the reduced 1D and 2D equations.  In Fig. 1, the numerical and variational 
densities are pretty close to each other, so are the results obtained from the 3D equation 
(\ref{gpe3d}), on the one hand, and  the ones obtained from the 1D and 2D equations
(\ref{gpe1d}) and (\ref{gpe2d}), on the other.

\begin{figure}[!t] 
\begin{center}  
\includegraphics[width=0.99\linewidth,clip]{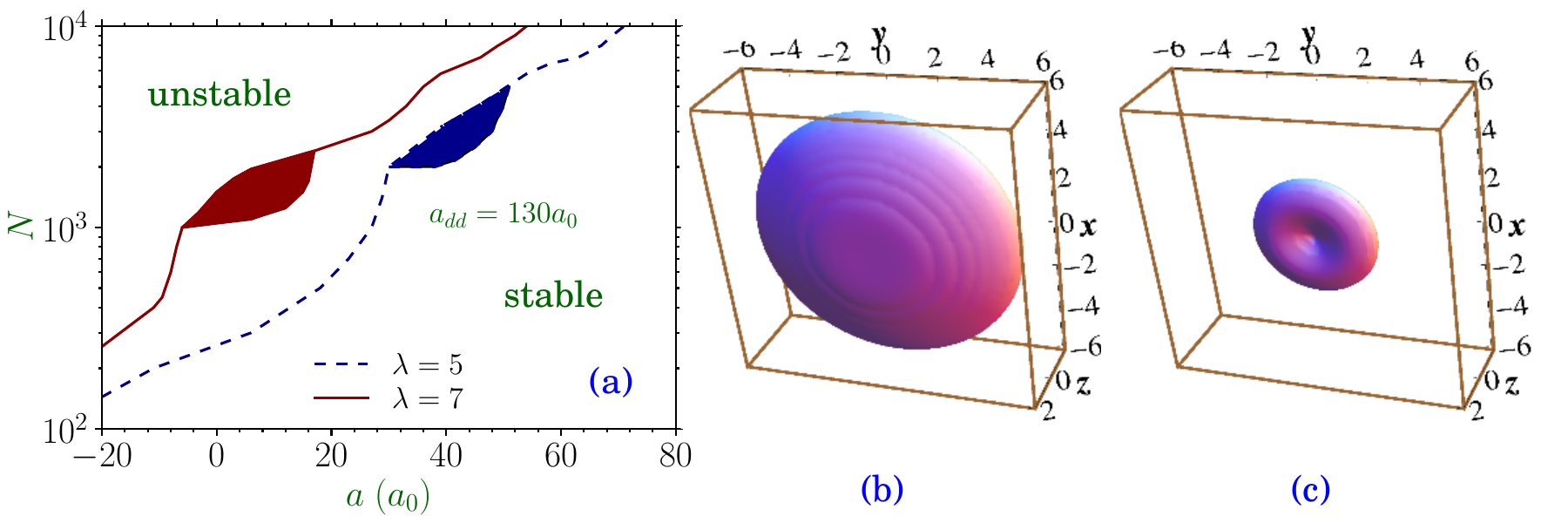} 
\end{center}
\caption{(a) The $N_{\mathrm{at}}-a$ stability phase plot for a $^{164}$Dy BEC with $a_{\mathrm{dd}}=130a_0$ 
in a disk-shaped 
trap with $\nu=\gamma=1$, $\lambda=5$ and 7 and harmonic oscillator length 
$l=1 $ $\mu$m. The 3D isodensity contour plot of density of a disk-shaped $^{164}$Dy BEC with $a_{\mathrm{dd}}=130a_0$ 
for $\nu=\gamma=1$, $\lambda=5$, $l=1 $ $\mu$m, $N_{\mathrm{at}}=3000$ and $a=40a_0$ for densities
$ \vert  \phi(x,y,z) \vert  ^2$ =
(b) 0.001  and (c) 0.027 on the contour. 
}
\label{fig2}
\end{figure}

A dipolar BEC is stable for the number of atoms $N_{\mathrm{at}}$ below a critical value \cite{Ronen2007}. 
Independent of trap parameters, such a BEC collapses as $N_{\mathrm{at}}$ crosses the critical  value. 
This can be studied by solving the 3D GP equation using imaginary-time propagation with a 
nonzero value of NSTP while the nonlinearities are slowly increased. 
In Fig. \ref{fig2} (a) 
we present the $N_{\mathrm{at}}-a$ stability phase plot  for a $^{164}$Dy BEC with $a_{\mathrm{dd}}=130a_0$
in the disk-shaped 
trap with $\nu=\gamma=1$, $\lambda=5$ and 7. The oscillator length is taken to be $l=1$ $\mu$m. 
The shaded area in these plots shows a metastable region where biconcave structure 
in 3D density appears. The metastable region corresponds to a local minimum in energy in contrast to a global minimum for a stable state. It has been established that this metastability is a manifestation 
of roton instability encountered by the system in the shaded region \cite{Ronen2007}. 
The biconcave structure in 3D density in a disk-shaped dipolar BEC is a direct consequence of 
dipolar interaction: the dipolar repulsion in the plane of the disk removes the atoms from the 
center to the peripheral region thus creating a biconcave shape in density. In Figs. \ref{fig2}
(b) and (c) we plot the 3D isodensity  contour of the condensate for $\lambda=5$ with parameters in the shaded region 
corresponding to metastability.
In Fig.  \ref{fig2} (b) the density on the contour is 0.001 whereas in 
Fig.  \ref{fig2} (c), it is  0.027. Only for a larger density on the contour the biconcave shape is visible. 
The biconcave shape predominates near the central region of the metastable dipolar BEC.

\begin{figure}[!t] 
\begin{center}  
\includegraphics[width=.47\linewidth,clip]{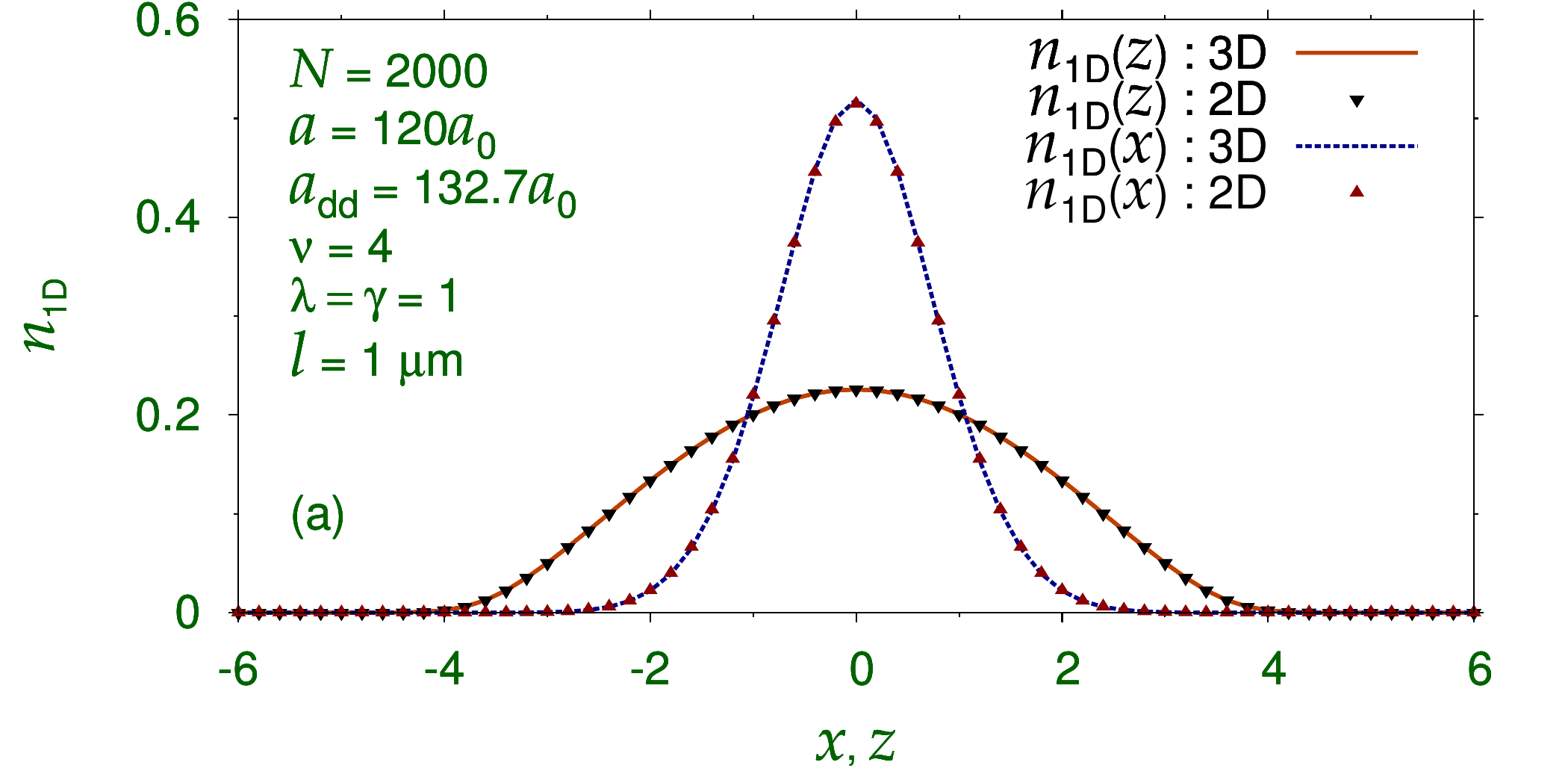} 
\includegraphics[width=.47\linewidth,clip]{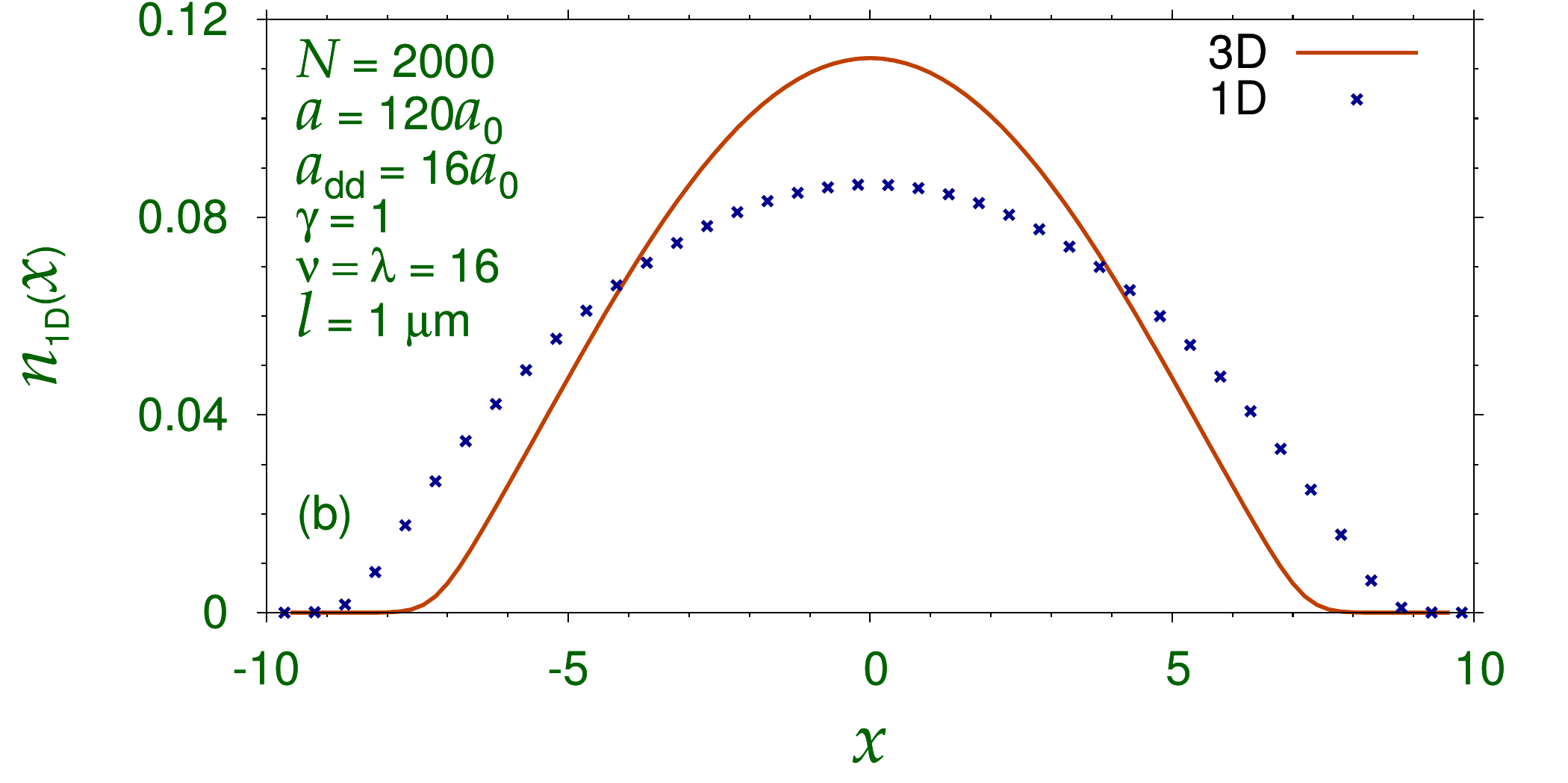}
\end{center}
\caption{
(a) Numerical   results for the  1D radial density 
$n_{1D}(x) =\int |\phi({\bf r})|^2 dy dz 
$ along $x$ axis
and $n_{1D}(z) =\int |\phi({\bf r})|^2 dx dy 
$ along $z$ axis
for $\lambda=\gamma=1, \nu =4$
 of a disk-shaped BEC of $N_{\mathrm{at}}=2000$ $^{164}$Dy atoms 
  obtained after integrating the 3D density from Eq. (\ref{gpe3d})   and   the 2D density from Eq. (\ref{gpe2dxz}) over the eliminated variables. 
(b) Numerical  results for the
1D axial density $n_{1D}(x)= $ along $x$ axis
for $\nu=\lambda=16, \gamma =1$
 of a cigar-shaped BEC of $N_{\mathrm{at}}=2000$ $^{52}$Cr
atoms obtained using the 
1D Eq.   (\ref{gpe1dx}) 
and that obtained after integrating the 3D density from Eq. (\ref{gpe3d}) over $z$ and $y$: 
$n_{1D}(x)= \int |\phi({\bf r})|^2 dz dy$.
In all cases $a=120a_0$  and (a) $a_{dd}= 132.7a_0$, (b)  $a_{dd}= 16a_0$.
}
\label{fig3}
\end{figure}

In figure \ref{fig1} we critically tested the reduced 1D and 2D equations (\ref{gpe1d}) and (\ref{gpe2d})
along the $z$ axis and in the $x-y$ plane, respectively, by comparing the different 1D densities from these equations with those obtained from a solution of the 3D equation (\ref{gpe3d}) as well as with the variational densities. Now we perform a similar test with the reduced 1D and 2D equations (\ref{gpe1dx}) and (\ref{gpe2dxz}) along the $x$ axis and in the $x-z$ plane, respectively. We consider a   BEC of 2000 atoms in a disk-shaped trap in the $x-z$ plane
 with $\lambda =\gamma=1$ and $\nu =4 $. Because of the strong trap in the $y$ direction, the resultant BEC is of quasi-2D shape in the $x-z$ plane without circular symmetry in that plane because of the anisotropic dipolar interaction.  The integrated linear density along the $x$ and $z$ axes as calculated from the 2D GP equation (\ref{gpe2dxz}) and the 3D GP equation (\ref{gpe3d})  are illustrated in figure \ref{fig3} (a).  Next we consider the BEC of 2000  atoms in a 
cigar-shaped trap along the $x$ axis with  $\nu=\lambda=16$ and $\gamma =1$. The integrated 
linear density along the $x$ axis in this case calculated from the 3D equation (\ref{gpe3d}) is compared with the same as calculate using the reduced 1D equation    (\ref{gpe1dx}) in figure \ref{fig3} (b). In both cases the densities calculated from the 3D GP equation are in reasonable agreement with those calculated using the reduced equations   (\ref{gpe2dxz}) and (\ref{gpe1dx}).
Another interesting feature emerges from figures \ref{fig1} and \ref{fig3}: the reduced 2D GP equations (\ref{gpe2d}) and (\ref{gpe2dxz}) with appropriate disk-shaped traps yield results for densities in better agreement with the 3D GP equation (\ref{gpe3d}) as compared to the 1D GP equations (\ref{gpe1d}) and (\ref{gpe1dx}) with appropriate cigar-shaped traps. This feature, also observed in non-dipolar BECs \cite{lsa}, is expected as the derivation of the 
reduced 1D equations involving two spatial integrations represent more drastic approximation compared to the  same of the reduced 2D equations involving one spatial integration.

\begin{figure}[!t] 
\begin{center}  
\includegraphics[width=.47\linewidth,clip]{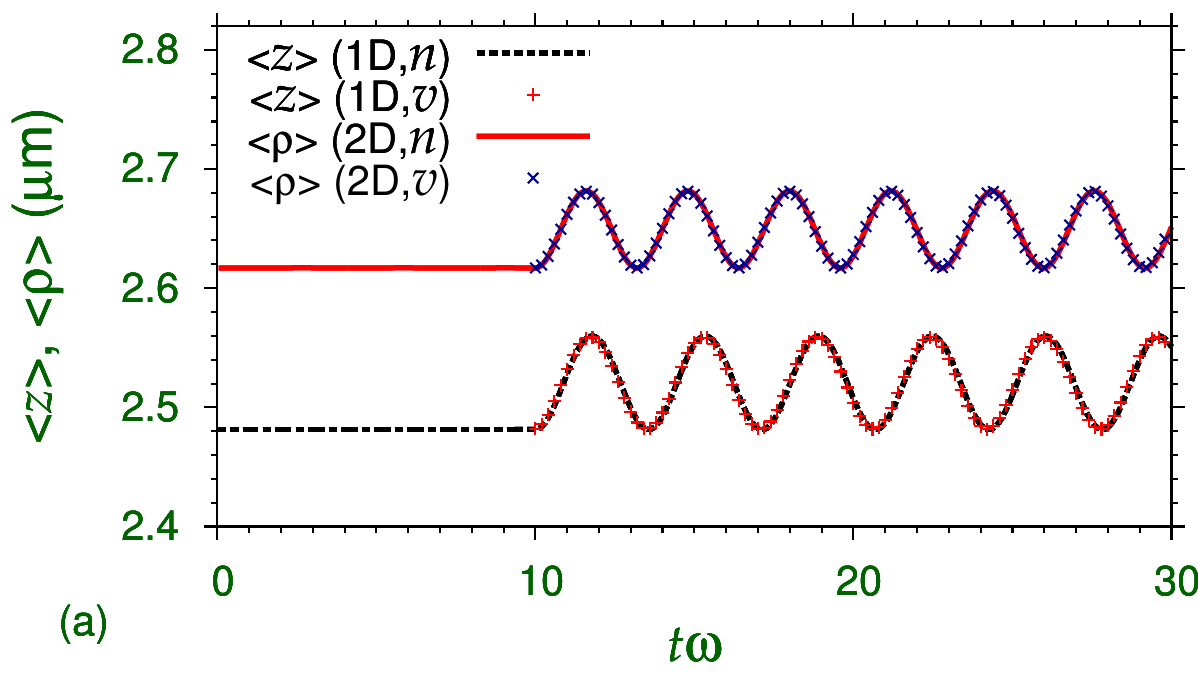} 
\includegraphics[width=.47\linewidth,clip]{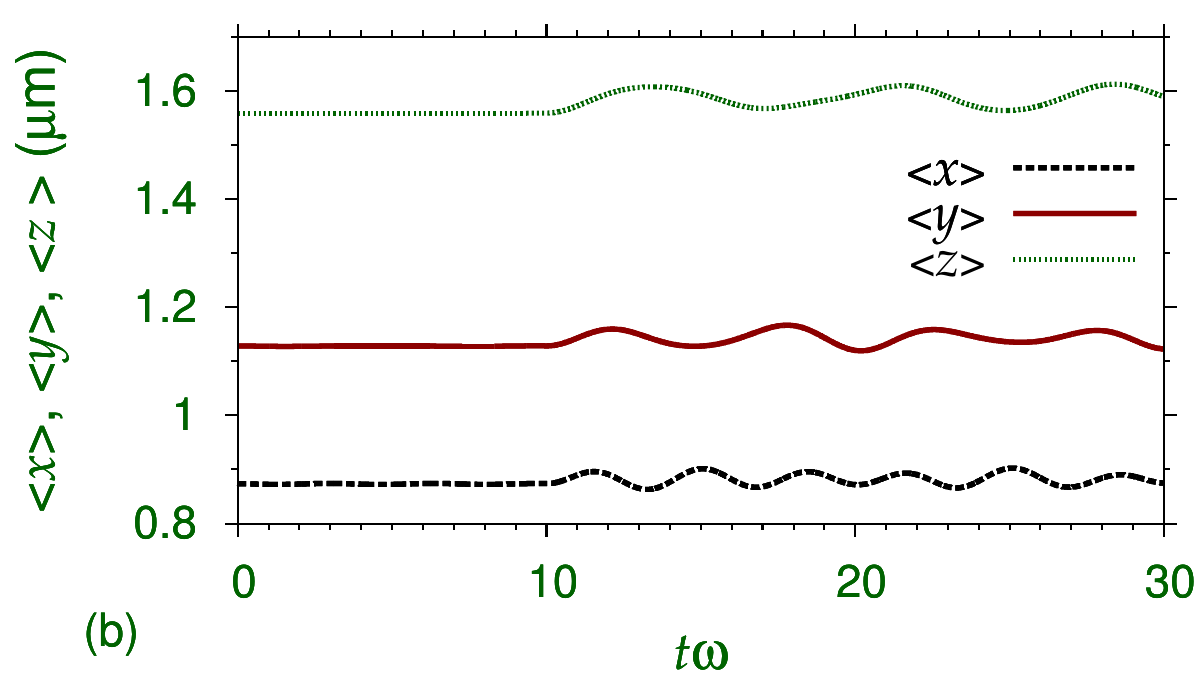}
\end{center}
\caption{(a) Numerical ($n$) and variational ($v$) results for  oscillation of rms sizes $\langle  z \rangle$ and  $\langle \rho  \rangle$ from  
the real-time simulation using Eq. (\ref{gpe1d})  in 1D and Eq. (\ref{gpe2d}) in   2D, respectively,  for $N_{\mathrm{at}}=10000$,  $a= 6$ nm, $a_{\mathrm{dd}}=16a_0$, $l=1$ $\mu$m, while $a$ and $a_{dd}$ were both multiplied by  1.05 after NPAS iterations at $t=10$. The wave function was first calculated by  imaginary-time routine with parameters   
$dx=0.025$, $dt=0.0001, \lambda=1, d_\rho =1 $, NPAS= $10^5, N = 2048$ in 1D, and   $dx = dy = 0.2, dt=0.001, \gamma=1,   d_z=1$ NPAS =  10$^4$, $NX = NY = 512$ in 2D. 
The results of the variational approximations in 1D and 2D as obtained from a numerical solution 
of Eqs. (\ref{vw1d})  and (\ref{vw2d}) are also shown.
(b) Numerical results for  oscillation of rms sizes $\langle x  \rangle, \langle y  \rangle$ and  $\langle z  \rangle$ from  
the real-time simulation in 3D using Eq. (\ref{gpe3d}),  for $N_{\mathrm{at}}=1000$,  $a= 110a_0$, $a_{\mathrm{dd}}=16a_0$, $l=1$ $\mu$m, $\gamma =1, \nu=1/\sqrt 2,
\lambda =1/2$,  $NX = NY = NZ = 128,
dx = dy = dz = 0.2$, and $dt = 0.002$ while $a$ and $a_{\mathrm{dd}}$ were both multiplied by  1.1 after NPAS iterations. 
In all cases the real-time   calculation was 
performed with NSTP = 0   reading the 3D density from the numerical solution of the imaginary-time program using the same 
parameters.}
\label{fig4}
\end{figure}

Now we report the dynamics of the dipolar BEC by  real-time propagation using the stationary 
state calculated by imaginary-time propagation. In Fig. \ref{fig4} (a) we show 
the oscillation of the rms sizes $\langle z\rangle$ and $\langle \rho \rangle$ from the  
reduced 1D and 2D GP equations (\ref{gpe1d}) and (\ref{gpe2d}), respectively. In Fig. \ref{fig4} (a) we consider 
$N_{\mathrm{at}}=10000, a_{\mathrm{dd}}=16a_0$ (appropriate for $^{52}$Cr), $a=6$ nm ($\approx 113a_0$) 
and oscillator length $l=1$ $\mu$m. In 1D, 
we take $dx = 0.025, dt =0.0001, \lambda =1, d_\rho =1$, number of space points $N = 2048$,   and in 2D, we take  $dx=dy= 0.2, dt =0.001, \gamma=\nu=1, 
d_z =1, NX = NY = 512. $in   real-time simulation the oscillation is started by multiplying the nonliniarities with the  factor
1.05. To impliment this, in real-time routine 
we take GPAR = GDPAR = 1.1 and also take NSTP = 0 to read the initial wave function.
In 1D and 2D we also present results of the Gaussian variational approximations after a numerical solution of Eqs. 
(\ref{vw1d}) and (\ref{vw2d}), respectively. The frequency of the resultant oscillations 
agree well with the numerical 1D and 2D calculations. However, slight adjustment of the 
  initial conditions, or initial values of width and its derivative, were necessary to get an agreement of the amplitude  of 
 oscillation obtained from variational approximation and numerical simulation. The initial values of width and its derivative are necessary to solve the variational Eqs. (\ref{vw1d}) and (\ref{vw2d}).
 In Fig. \ref{fig4} (b) we illustrate the oscillation of the rms sizes  
$\langle x\rangle$,  $\langle y\rangle$, and  $\langle z\rangle$ in 3D using Eq. (\ref{gpe3d}), where we perform real-time simulation using the bound state obtained by imaginary-time simulation as the initial state. The parameters used are
  $N_{\mathrm{at}}=1000, a=110a_0, 
a_{\mathrm{dd}}=16a_0, \gamma=1, \nu =1/\sqrt 2, \lambda =1/2, l = 1$ $\mu$m, $NX = NY = NZ = 128,
dx = dy = dz = 0.2,  dt = 0.002$ in both real- and  imaginary-time simulations. In addition, 
in   real-time simulation the oscillation is started by multiplying the nonliniarities with the  factor
1.1. To impliment this, in real-time routine 
we take GPAR = GDPAR = 1.1 and also take NSTP = 0 to read the initial wave function.

\section{Summary}

\label{summary}

We have presented useful numerical programs in Fortran and C for solving the dipolar GP equation including the contact interaction
in 1D, 2D, 3D. Two sets of programs are provided. The imaginary-time programs are appropriate 
for solving the stationary problems, while the real-time codes can be used for studying non-stationary dynamics. 
The programs are developed in Cartesian coordinates. We have compared the results of numerical 
calculations for statics and dynamics of dipolar BECs with those of Gaussian variational approximation, Thomas-Fermi  approximation, and numerical calculations of other authors, where 
possible. 

\section*{Acknowledgements}
The authors thank Drs. Weizhu Bao, Doerte Blume, Hiroki Saito,
and Luis Santos for helpful comments on numerical calculations.
RKK acknowledges support from the TWAS (Third World Academy
of Science, Trieste, Italy) - CNPq (Brazil) project Fr 3240256079,
DST (India) - DAAD (Germany) project SR/S2/HEP-03/2009, PM
from CSIR (Project 03(1186)/10/EMR-II, (India), DST-DAAD (Indo-
German, project INT/FRG/DAAD/P-220/2012), SKA from the CNPq
project 303280/2014-0 (Brazil) and FAPESP project 2012/00451-0
(Brazil), LEYS from the FAPESP project 2012/21871-7 (Brazil). DV
and AB acknowledge support by the Ministry of Education, Science
and Technological Development of the Republic of Serbia under
project ON171017 and by DAAD (Germany) under project NAIDBEC,
and by the European Commission under EU FP7 projects
PRACE-3IP and EGI-InSPIRE.


\begin{thebibliography}{100}



\bibitem{Dalfovo1999}
F.~Dalfovo, S.~Giorgini, L.~P. Pitaevskii, S.~Stringari, Theory of
  Bose-Einstein condensation in trapped gases, Rev. Mod. Phys. 71 (1999)
  463-512;

A.J. Leggett, Bose-Einstein condensation in the alkali gases: Some fundamental concepts, Rev. Mod. Phys. 73 (2001) 307–356;

 L. Pitaevskii, S. Stringari, Bose-Einstein Condensation, Clarendon Press, Oxford and New York, 2003;

C.J. Pethick, H. Smith, Bose-Einstein Condensation in Dilute Gases, Cambridge University Press, Cambridge, 2002.


\bibitem{mu-ad}S. K. Adhikari, P. Muruganandam, Bose-Einstein condensation dynamics from the numerical solution of the Gross-Pitaevskii equation, J. Phys. B 35 (2002) 2831;



P. Muruganandam, S. K. Adhikari, 
Bose-Einstein condensation dynamics in three dimensions by the pseudospectral and finite-difference methods, 
J. Phys. B 36 (2003) 2501;

S.K. Adhikari, Numerical study of the spherically symmetric Gross-Pitaevskii
equation in two space dimensions, Phys. Rev. E 62 (2000) 2937–2944;

S.K. Adhikari, Numerical solution of the two-dimensional Gross-Pitaevskii
equation for trapped interacting atoms, Phys. Lett. A 265 (2000) 91–96.
\bibitem{CPC1}P. Muruganandam, S. K. Adhikari, 
Fortran programs for the time-dependent Gross-Pitaevskii equation in a fully anisotropic trap,
Comput. Phys. Commun. 180
 (2009)   1888-1912.


\bibitem{CPC2}     D. Vudragovi\'c,
     I. Vidanovi\'c,
     A. Bala\v{z}, 
     P. Muruganandam,
 S. K. Adhikari, C programs for solving the time-dependent 
Gross-Pitaevskii equation in a fully anisotropic trap, 
Comput. Phys. Commun. 
183 (2012) 2021-2025.

\bibitem{use}
W. Wen, C. Zhao, X. Ma, 
Dark-soliton dynamics and snake instability in superfluid Fermi gases trapped by an anisotropic harmonic potential
Phys. Rev. A  88 (2013)  063621;   

  K.-T. Xi, J. Li, D.-N. Shi, 
Localization of a two-component Bose-Einstein condensate in a two-dimensional bichromatic optical lattice,
Physica B-Cond. Mat.   436 (2014)   149-156;   


Y.-S. Wang, Z.-Y. Li, Z.-W. Zhou, X.-F. Diao, 
Symmetry breaking and a dynamical property of a dipolar Bose-Einstein condensate in a double-well potential,
Phys. Lett. A 378 (2014) 48-52;


E. J. M. Madarassy, V. T.  Toth, 
Numerical simulation code for self-gravitating Bose-Einstein condensates,
Comput. Phys. Commun. 184 (2013) 1339-1343;

R. M. Caplan,
 NLSEmagic: Nonlinear Schr\"odinger equation multi-dimensional Matlab-based GPU-accelerated integrators using compact high-order schemes,
Comput. Phys. Commun. 184 (2013) 1250-1271;

 







I. Vidanovic, A. Bala\v{z}, H. Al-Jibbouri, A. Pelster, Nonlinear Bose-Einstein-condensate dynamics induced by a harmonic
modulation of the s-wave scattering length, Phys. Rev. A 84 (2011) 013618;

Y. Cai, H. Wang, 
Analysis and computation for ground state solutions of Bose-Fermi mixtures at zero temperature,
Siam J. App. Math. 73 (2013) 757-779;



A. Bala\v{z}, R. Paun, A. I. Nicolin, S. Balasubramanian,  R. Ramaswamy, 
 Faraday waves in collisionally inhomogeneous Bose-Einstein condensates,
Phys. Rev. A 89 (2014) 023609;








H. Al-Jibbouri, A. Pelster, 
Breakdown of the Kohn theorem near a Feshbach resonance in a magnetic trap, Phys. Rev. A 88 (2013) 033621;



E. Yomba, G.-A. Zakeri,  
Solitons in a generalized space- and time-variable coefficients nonlinear Schr\"odinger equation with higher-order terms,
Phys. Lett. A 377 (2013) 2995-3004;


X. Antoine, W. Bao, C. Besse, 
Computational methods for the dynamics of the nonlinear Schr\"odinger/Gross-Pitaevskii equations,
Comput. Phys. Commun. 184 (2013) 2621-2633;

N. Murray {\it et al.},
    Probing the circulation of ring-shaped Bose-Einstein condensates,
Phys. Rev. A 88 (2013) 053615;





A. Khan, P. K. Panigrahi, 
 Bell solitons in ultra-cold atomic Fermi gas, 
J. Phys. B 46 (2013) 115302;




E. Yomba, G.-A. Zakeri,
Exact solutions in nonlinearly coupled cubic-quintic complex Ginzburg-Landau equations,
Phys. Lett. A 377 (2013) 148-157;


W. Bao, Q. Tang, Z. Xu, 
Numerical methods and comparison for computing dark and bright solitons in the nonlinear Schr\"odinger equation,
J. Comput. Phys. 235 (2013) 423-445;


W. Wen, H.-J. Li,
 Interference between two superfluid Fermi gases, J. Phys. B
46 (2013) 035302;


H. Zheng, Y. Hao, Q. Gu,
Dynamics of double-well Bose-Einstein condensates subject to external Gaussian white noise,
J. Phys. B 46 (2013) 065301;

Y. Wang, F.-D. Zong, F.-B. Li,
Three-dimensional Bose-Einstein condensate vortex solitons under optical lattice and harmonic confinements,
Chinese Phys. B. 22 (2013) 030315;


P.-G. Yan, S.-T. Ji,  X.-S. Liu,
Symmetry breaking and tunneling dynamics of F=1 spinor Bose-Einstein condensates in a triple-well potential,
Phys. Lett. A 377 (2013) 878-884;



R. R. Sakhel, A. R. Sakhel, H. B.  Ghassib, 
Nonequilibrium Dynamics of a Bose-Einstein Condensate Excited by a Red Laser Inside a Power-Law Trap with Hard Walls, J. Low Temp. Phys
173 (2013) 177-206;

A. Trichet, E. Durupt, F. M\'edard, S. Datta, A. Minguzzi, and M. Richard,
Long-range correlations in a 97$\%$ excitonic one-dimensional polariton condensate, Phys. Rev. B 88 (2013) 121407;






X. Yue {\it et al.},
Observation of diffraction phases in matter-wave scattering,
Phys. Rev. A 88 (2013) 013603;

S. Prabhakar {\it et al.},
Annihilation of vortex dipoles in an oblate Bose-Einstein condensate,
J. Phys. B 46 (2013) 125302;




Z. Marojevic, E. Goeklue, C. Laemmerzahl, 
Energy eigenfunctions of the 1D Gross-Pitaevskii equation,
Comput. Phys. Commun. 184 (2013) 1920-1930;


T. Mithun, K. Porsezian, B. Dey, 
Vortex dynamics in cubic-quintic Bose-Einstein condensates,
Phys. Rev. E 88 (2013) 012904;


M. Edwards, M.  Krygier, H.  Seddiqi, B. Benton, and C. W. Clark,
Approximate mean-field equations of motion for quasi-two-dimensional Bose-Einstein-condensate systems,
Phys. Rev. E 86 (2012) 056710;



J. Li, F.-D. Zong, C.-S. Song, Y. Wang, and F.-B. Li, Dynamics of analytical three-dimensional solutions in Bose-Einstein condensates with time-dependent gain and potential,
Phys. Rev. E 85 (2012) 036607;




 



P. Verma, A. B. Bhattacherjee, M. Mohan, Oscillations in a parametrically excited Bose-Einstein condensate in combined harmonic and optical lattice trap, Central Eur. J. Phys. 10 (2012) 335-341;

E. R. F. Ramos, F. E. A. dos Santos, M. A. Caracanhas, and V. S. Bagnato, Coupling collective modes in a trapped superfluid, Phys. Rev. A 85 (2012) 033608; 
 



  W. B. Cardoso, A. T. Avelar, D. Bazeia, Modulation of localized solutions in a system of two coupled nonlinear Schr\"odinger equations, 
Phys. Rev. E 86 (2012) 027601;


P.-G. Yan, Y.-S. Wang, S.-T Ji, X.-S. Liu, Symmetry breaking of a Bose-Fermi mixture in a triple-well potential, Phys. Lett. A 376 (2012) 3141-3145;








H. L. Zheng, Y. J. Hao, Q. Gu,
 Dissipation effect in the double-well Bose-Einstein condensate,
Eur. Phys. J. D 66 (2012) 320;



P. Verma, A. B. Bhattacherjee, M. Mohan, 
 Oscillations in a parametrically excited Bose-Einstein condensate in combined harmonic and optical lattice trap, Central Eur. J. Phys. 
10 (2012) 335-341;
 

 

 W. B. Cardoso, A. T. Avelar, D. Bazeia, One-dimensional reduction of the three-dimenstional Gross-Pitaevskii equation
with two- and three-body interactions, Phys. Rev. E 83 (2011) 036604;


 R. R. Sakhel, A. R. Sakhel, H. B. Ghassib, Self-interfering matter-wave patterns generated by a moving laser obstacle
in a two-dimensional Bose-Einstein condensate inside a power trap cut-off by box potential boundaries, Phys. Rev. A 84
(2011) 033634;

 A. Bala\v{z}, A. I. Nicolin, Faraday waves in binary nonmiscible Bose-Einstein condensates, Phys. Rev. A 85 (2012) 023613;
 

 S. Yang, M. Al-Amri, J. Evers, M. S. Zubairy, Controllable optical switch using a Bose-Einstein condensate in an optical
cavity, Phys. Rev. A 83 (2011) 053821;



 Z. Sun, W. Yang, An exact short-time solver for the time-dependent Schr\"odinger equation, J. Chem. Phys. 134 (2011)
041101;





 G. K. Chaudhary, R. Ramakumar, Collapse dynamics of a (176)Yb-(174)Yb Bose-Einstein condensate, Phys. Rev. A 81
(2010) 063603;



 S. Gautam, D. Angom, Rayleigh-Taylor instability in binary condensates, Phys. Rev. A 81 (2010) 053616;

S. Gautam, D. Angom, Ground state geometry of binary condensates in axisymmetric traps, J. Phys. B 43 (2010) 095302;



 G. Mazzarella, L. Salasnich, Collapse of triaxial bright solitons in atomic Bose-Einstein condensates, Phys. Lett. A 373
(2009) 4434-4437;


 


\bibitem{cr}
T.~Koch, T.~Lahaye, Fr\"ohlich, A.~Griesmaier, T.~Pfau, {S}tabilization of a
  purely dipolar quantum gas against collapse, Nature Phys. 4 (2008) 218-222.




\bibitem{dy}M. Lu, N. Q. Burdick, Seo Ho Youn, B. L. Lev, 
Strongly Dipolar Bose-Einstein Condensate of Dysprosium,
Phys.
Rev. Lett. 107 (2011) 190401.

  



\bibitem{er}K. Aikawa {\it et al.}, 
Bose-Einstein Condensation of Erbium,
Phys. Rev. Lett. 108 (2012) 210401.



\bibitem{Bao2010}
W.~Bao, Y.~Cai, H.~Wang, Efficient numerical methods for computing ground
  states and dynamics of dipolar {B}ose-{E}instein condensates, J. Comput.
Phys. {229} (2010) 7874-7892.



\bibitem{Yi2001}
S.~Yi, L.~You, Trapped condensates of atoms with dipole interactions, Phys.
  Rev. A 63 (2001) 053607;


 Quantum Phases of Dipolar Spinor Condensates,
S. Yi, L. You, and H. Pu,
Phys. Rev. Lett. 93 (2004) 040403.



\bibitem{Goral2002a}
K.~G\'oral, L.~Santos, Ground state and elementary excitations of single and
  binary {B}ose-{E}instein condensates of trapped dipolar gases, Phys. Rev. A
  66 (2002) 023613.




\bibitem{dip3}S. Ronen, D. C. E. Bortolotti, J. L. Bohn, Bogoliubov modes of a dipolar condensate in a cylindrical trap, Phys. Rev.  A 74
(2006) 013623. 



\bibitem{Blakie}P. B. Blakie, C. Ticknor, A. S. Bradley, A. M. Martin, M. J. Davis, Y. Kawaguchi, 
Numerical method for evolving the dipolar projected Gross-Pitaevskii equation,
Phys. Rev. E 80 (2009) 016703.  

\bibitem{dip5} T. Lahaye, J. Metz, B. Fröhlich, T. Koch, M. Meister, A. Griesmaier, T. Pfau, H. Saito, Y. Kawaguchi, and M. Ueda,  $d$-Wave Collapse and Explosion of a Dipolar Bose-Einstein Condensate, 
Phys. Rev. Lett. 101 (2008) 080401. 




\bibitem{Parker}N. G. Parker, C. Ticknor, A. M. Martin, D. H. J. O'Dell, Structure formation during the collapse of a dipolar atomic Bose-Einstein condensate, Phys. Rev. A 79 (2009)  013617;



C.~Ticknor, N.~G. Parker, A.~Melatos, S.~L. Cornish, D.~H.~J. O'Dell, A.~M.
  Martin, Collapse times of dipolar {B}ose-{E}instein condensate, Phys. Rev. A
78 (2008)  061607.
 



\bibitem{Blume}
M. ~Asad-uz-Zaman, D. ~Blume, Aligned dipolar bose-einstein condensate in a
  double-well potential: From cigar shaped to pancake shaped, Phys. Rev. A 80
  (2009) 053622.

\bibitem{lsa}L. Salasnich, A. Parola, and L. Reatto, 
Effective wave equations for the dynamics of cigar-shaped and disk-shaped Bose condensates,
Phys. Rev. A 65   (2002) 043614.


\bibitem{Santos2000}
L.~Santos, V.~Shylapnikov, G, P.~Zoller, M.~Lewenstein, {B}ose-{E}instein
  condensation in trapped dipolar gases, Phys. Rev. Lett. 85 (2000) 1791-1794.


\bibitem{prog} T Lahaye, C Menotti, L Santos, M Lewenstein, T Pfau,
 The physics of dipolar bosonic quantum gases, Rep. Prog. Phys. 72 (2009)
126401.


\bibitem{gv}
V. M. P\'erez-García, H. Michinel, J. I. Cirac, M. Lewenstein, and P. Zoller 
Phys. Rev. A
{56} (1997) 1424-1432. 




\bibitem{joptb}
S. Giovanazzi, A. G\"orlitz, T. Pfau, 
Ballistic expansion of a dipolar condensate, J. Opt. B {5} (2003) S208-S211.


\bibitem{laserphysics}P. Muruganandam, S. K. Adhikari,
Numerical and variational solutions of the dipolar Gross-Pitaevskii equation in reduced dimensions,
Laser Phys. {22} (2012) 813-820.


\bibitem{ODell2004}
D.~O'Dell, S.~Giovanazzi, C.~Eberlein, {E}xact {H}ydrodynamics of a dipolar
  {B}ose-{E}instein condensate, Phys. Rev. Lett. 92 (2004) 250401.

\bibitem{Eberlein2005}
C.~Eberlein, S.~Giovanazzi, D.~H.~J. O'Dell, {E}xact solution of the
  {T}homas-{F}ermi equation for a trapped {B}ose-{E}instein condensate with
  dipole-dipole interactions, Phys. Rev. A 71 (2005) 033618.


\bibitem{Parker2008}
N.~G. Parker, D.~H.~J. O'Dell, {T}homas-{F}ermi versus one- and two-dimensional
  regimes of a trapped dipolar {B}ose-{E}instein condensate, Phys. Rev. A. 78
  (2008) 41601(R).



\bibitem{santos1d}  S. Sinha, L. Santos, 
Cold dipolar gases in quasi-one-dimensional geometries,
Phys. Rev. Lett. 99 (2007) 140406;

 F. Deuretzbacher, J. C. Cremon,  S. M. Reimann, Ground-state properties of few dipolar bosons in a quasi-one-dimensional harmonic trap,
Phys. Rev. A 81 (2010) 063616.


\bibitem{1d-dell}S. Giovanazzi,  D. H. J. O'Dell, Eur. Phys. J. D  31 (2004) 439-445.



\bibitem{santos2d}U. R. Fischer, Stability of quasi-two-dimensional Bose-Einstein condensates with dominant dipole-dipole interactions, Phys. Rev. A 73 (2006) 031602;

 P. Pedri,  L. Santos, Two-Dimensional Bright Solitons in Dipolar Bose-Einstein Condensates, Phys. Rev. Lett. 95 (2005) 200404.


\bibitem{Blume2}M. Asad-uz-Zaman, D. Blume, private communication (2010). 



 \bibitem{afta}A. Aftalion, Q. Du, Vortices in a rotating Bose–Einstein condensate: Critical angular velocities and energy diagrams in the Thomas-Fermi regime, Phys. Rev. A 64 (2001)
063603.

\bibitem{Ronen2007}
S.~Ronen, D.~C.~E. Bortolotti, J. L. Bohn, {R}adial and {A}ngular {R}otons in
  {T}rapped {D}ipolar {G}ases, Phys. Rev. Lett. 98 (2007) 030406; 
 
R. M. Wilson, S. Ronen, J. L. Bohn,  H. Pu,
 Manifestations of the Roton Mode in Dipolar Bose-Einstein Condensates,
Phys. Rev. Lett. 100, (2008) 245302;

L. Santos, G. V. Shlyapnikov,  M. Lewenstein,
 Roton-Maxon Spectrum and Stability of Trapped Dipolar Bose-Einstein Condensates,
Phys. Rev. Lett. 90 (2003) 250403;

M. Asad-uz-Zaman, D. Blume,
 Modification of roton instability due to the presence of a second dipolar Bose-Einstein condensate,
Phys. Rev. A 83 (2011) 033616. 


\end{thebibliography}
\end{document}